\documentclass[letterpaper,floatfix,aps,pra,amsmath,amssymb,twocolumn,superscriptaddress,
showpacs]{revtex4-1}

\usepackage{epsfig}
\usepackage{verbatim}
\usepackage{color}
\usepackage{epsf}
\usepackage{array}
\usepackage[pdfauthor={Hosseinkhani, Riwar, Schoelkopf, Glazman, Catelani},
            pdftitle={Optimal configurations for normal-metal traps in transmon qubits}]{hyperref}

\hypersetup{pdfstartview={XYZ null null 1.05}}

\newcommand{\be}{\begin{equation}}
\newcommand{\ee}{\end{equation}}
\newcommand{\bea}{\begin{eqnarray}}
\newcommand{\eea}{\end{eqnarray}}

\newcommand{\eref}[1]{Eq.~(\ref{#1})}
\newcommand{\esref}[1]{Eqs.~(\ref{#1})}
\newcommand{\rref}[1]{(\ref{#1})}

\newcommand{\ocite}[1]{Ref.~\cite{#1}}

\newcommand{\qp}{\mathrm{qp}}

\usepackage{epstopdf}
\begin{document}

\title{Optimal configurations for normal-metal traps in transmon qubits}

\author{A. Hosseinkhani}

\affiliation{JARA-Institute for Quantum Information (PGI-11), Forschungszentrum J\"ulich, 52425 J\"ulich, Germany}
\affiliation{JARA-Institute for Quantum Information, RWTH Aachen University, D-52056 Aachen, Germany}

\author{R.-P. Riwar}

\affiliation{JARA-Institute for Quantum Information (PGI-11), Forschungszentrum J\"ulich, 52425 J\"ulich, Germany}
\affiliation{Departments of Physics and Applied Physics, Yale University, New Haven, CT 06520, USA}

\author{R. J. Schoelkopf}

\affiliation{Departments of Physics and Applied Physics, Yale University, New Haven, CT 06520, USA}

\author{L. I. Glazman}

\affiliation{Departments of Physics and Applied Physics, Yale University, New Haven, CT 06520, USA}

\author{G. Catelani}

\affiliation{JARA-Institute for Quantum Information (PGI-11), Forschungszentrum J\"ulich, 52425 J\"ulich, Germany}

\begin{abstract}
Controlling quasiparticle dynamics can improve the performance of superconducting devices. For example, it has been demonstrated effective in increasing lifetime and stability of superconducting qubits. Here we study how to optimize the placement of normal-metal traps in transmon-type qubits.
When the trap size increases beyond a certain characteristic length, the details of the geometry and trap position, and even the number of traps, become important.
We discuss for some experimentally relevant examples how to shorten the decay time of the excess quasiparticle density.
Moreover, we show that a trap in the vicinity of a Josephson junction can significantly reduce the steady-state quasiparticle density near that junction, thus suppressing the quasiparticle-induced relaxation rate of the qubit.
Such a trap also reduces the impact of fluctuations in the generation rate of quasiparticles, rendering the qubit more stable.
\end{abstract}

\date{\today}

\maketitle

\section{Introduction}

For a successful execution of quantum gates, the participating qubits must have coherence times exceeding the gate operation time. Moreover, the longer the coherence time, the less resources are needed for the implementation of error correction. Thus, it is important to understand and control decoherence mechanisms.
Over the past several years it has been firmly established, both theoretically \cite{lutchyn,prl,leppa} and experimentally \cite{shaw,martinis,3dtr,riste,nature}, that quasiparticles are detrimental to superconducting qubits based on Josephson junctions. For example, tunneling of quasiparticles through a transmon junction leads to a relaxation rate proportional to their density. At the low temperatures the qubits are operated, the number of quasiparticles in typical devices should be negligible in thermal equilibrium; however, the measured quasiparticle density is several orders of magnitude larger than expected, indicating that quasiparticle generation mechanisms of unknown origin are maintaining a large non-equilibrium quasiparticle population. While in general it is not possible to control the poorly understood generation processes, trapping quasiparticles away from the junctions offers a way to limit their unwanted effects. In this paper we build on a recently developed model~\cite{Riwar} for trapping by normal-metal islands to explore ways to optimize the traps performance.

A number of approaches to trap quasiparticles have been explored, from gap engineering~\cite{cpt,sun} to trapping by
vortices~\cite{ullom,plourde,wang,pekola2} and normal-metal islands~\cite{court,raja1,raja2,pekolaprb}. In all cases, trapping is made possible by energy relaxation of excitations inside the trap, since an excitation with energy below the gap $\Delta$ of the device's superconductor $S$ cannot return to $S$. For normal-metal traps in tunnel contact with $S$, the interplay between tunneling and relaxation was studied both theoretically and experimentally in \ocite{Riwar}, with the measurements indicating that relaxation is the bottleneck limiting the trapping rate.
The considerations in \ocite{Riwar} were mostly limited to the decay dynamics of excess quasiparticles in a model of a point-like trap placed in a (quasi-)one-dimensional wire; in contrast, here we examine finite-size traps in realistic qubit geometries. We focus on a qubit design, the coplanar gap capacitor transmon of Ref.~\cite{wang}, whose coherence was shown to be limited by quasiparticles. Similarly, the relaxation time of a fluxonium biased close to (but not at) the half flux quantum~\cite{nature} and that of a flux qubit~\cite{gustavsson} were found to be limited by quasiparticles. In other experiments with transmons (in a different design with large pads rather than gap capacitor plus small pads), quasiparticles were not the limiting factor, but there was evidence that they will become relevant if the coherence time is to be increased by another order of magnitude~\cite{riste}. In addition to this kind of qubits, quasiparticles are detrimental to both the so-called Andreev level qubit~\cite{shumeiko,saclay} and potentially to future devices based on Majorana states~\cite{loss} -- in both cases, coherence is destroyed by a single quasiparticle changing the parity of the state. Quasiparticles are also a source of errors in charge pumps comprising superconducting elements, which hampers their use in metrological applications -- indeed, traps have been used to increase the pump accuracy~\cite{pekolarmp}. Therefore, the analysis presented here for a particular case can potentially indicate the way for engineering better trap configurations in a variety of systems, since it is based on a phenomenological diffusion equation which is expected to hold quite generally.

In this paper we show that use of quasi-1D geometries facilitates trapping with small traps, and that their positions can be optimized.
We consider three ways in which normal-metal traps may improve qubit performance. First, we note that events which generate a large number of quasiparticles render the qubit inoperable so long as the excess quasiparticles are not eliminated; here we find the parameters and placement of traps that enhance the relaxation rate of the excess density.
Second, in addition to the dynamics of the excess density, we study the quasiparticles steady-state density in the presence of a generic generation mechanism with a rate determined by experiments~\cite{wang,vool}; we find that a trap placed near a junction can suppress the quasiparticle density at that junction, potentially leading to a longer $T_1$ relaxation time for the qubit. Third, we consider the effect of fluctuations in the generation rate: they lead to the fluctuation in the
quasiparticle density near the junction and, associated with it, to variations of a qubit $T_1$~\cite{vool,gustavsson}; placing a trap up to a certain distance from the junction can reduce the density fluctuations and hence make the qubit more stable.

The paper is organized as follows: in Sec.~\ref{sec:model} we summarize the model of \ocite{Riwar} to establish our notation and for the paper to be self-contained. In Sec.~\ref{sec:optimal} we consider a realistic qubit geometry, namely the coplanar gap capacitor transmon of Refs.~\cite{wang,Riwar};
we give analytical arguments for trap configurations leading to faster relaxation rates of the excess density -- see \eref{Lopt} for the single trap case and \esref{eq_multi_trap_decay_rate} and \rref{Nopt} for the multi-trap one -- and complement those with numerical calculations whose outcomes are summarized in Figs.~\ref{fig:devb1} to \ref{fig:dp}. In Sec.~\ref{sec:steady} we turn our attention to the steady-state density at the junction and its fluctuations; both can be suppressed by appropriately placed traps, but while the steady-state density always increases monotonically with trap-junction distance [\eref{xqpj2}], we find for a strong trap a non-monotonic behavior of fluctuations [\eref{Dxqp2s}] and hence an optimal trap position. We summarize our findings in Sec.~\ref{sec:summary}. A number of Appendices complement the main text: in Appendix~\ref{app:vortices} we compare trapping by normal-metal traps with that due to vortices; in Appendix~\ref{appendix_finite_size_trap} we present some mathematical details for the case of a single, finite-size trap, and Appendix~\ref{app:mode_degeneracy} addresses the question of the experimental observability of the slowest decay rate of the excess density; Appendices~\ref{app:Effective-length-pad} and \ref{app:Effective-length-wings} contain details about the mapping of a realistic qubit design into a 1D wire; finally Appendix~\ref{app:Xmon} considers traps in the Xmon qubit geometry.

\section{The effective trapping rate model}
\label{sec:model}

Superconducting qubits are generically fabricated by depositing thin superconducting films over an insulating substrate; we therefore consider quasiparticles as diffusing in a two-dimensional region. Normal-metal ($N$) traps are also thin films deposited on top of the superconductor $S$, and we assume that an insulating barrier of low transparency is present between $N$ and $S$.
Then the dynamics of the (normalized) quasiparticle density in the superconductor $x_\qp$
is captured by a generalized diffusion equation
\begin{equation}\label{eq_diff_eq}
\dot{x}_{\text{qp}}=D_{\text{qp}}\vec{\nabla}^{2}x_{\text{qp}}-\mathcal{A}\left(\vec{x}\right)\Gamma_{\text{eff}}x_{\text{qp}}-s_{\text{b}}x_{\text{qp}}+g
\end{equation}
(see \ocite{Riwar} for theoretical justification and experimental validation of this effective model).
The quasiparticle density $x_{\text{qp}}$ is normalized by the Cooper pair density $\nu_0\Delta$ and is therefore a dimensionless quantity; it is a function of time $t$ (the dot denotes the time derivative) and position $\vec{x}$ in the $x$-$y$ plane (here $\nu_0$ is the density of states at the Fermi level).
The phenomenological diffusion
constant $D_{\text{qp}}$  is proportional to the normal-state diffusion
constant for electrons in the superconductor, $D_{S}$. The term proportional to the effective trapping rate $\Gamma_\mathrm{eff}$ accounts for trapping of quasiparticles by the normal metal and is discussed in more detail below. The function $\mathcal{A}(\vec{x})$ is unity when $\vec{x}$ is within the $S$-$N$ contact region and zero otherwise -- in other words, this function localizes the trap to only a part of the whole device, rendering the system inhomogeneous, whereas the other coefficients in \eref{eq_diff_eq} are constant throughout the superconductor. The rate $g$ describes the generation of quasiparticles. Finally, the possibility that other mechanisms can trap quasiparticles in the bulk of the superconductor is captured by the term proportional to the background trapping rate $s_\mathrm{b}$; we set $s_b=0$ in this paper, as its effect is negligible~\cite{noterecomb}.

The effective trapping rate $\Gamma_{\text{eff}}$ incorporates the interplay between tunneling from $S$ to $N$ at rate $\Gamma_\textrm{tr}$, relaxation in $N$ with rate $\Gamma_\textrm{r}$, and escape from $N$ to $S$ with rate $\Gamma_\textrm{esc}$.
For a quasiparticle distribution at an (effective) temperature $T\ll\Delta$, we can distinguish two limiting cases.
For fast relaxation, $\Gamma_{\text{r}}\gg\left(\Delta/T\right)^{1/2}\Gamma_{\text{esc}}$,
the effective trapping rate is determined by the $S$ to $N$ tunneling rate, $\Gamma_{\text{eff}}\approx\Gamma_{\text{tr}}$: since quasiparticles loose energy quickly upon entering into $N$, they cannot tunnel back to $S$.
For slow relaxation, $\Gamma_{\text{r}}\ll \left(\Delta/T\right)^{1/2}\Gamma_{\text{esc}}$,
on the other hand, we have $\Gamma_{\text{eff}}\approx\left(2T/\pi\Delta\right)^{1/2}\Gamma_{\text{tr}}\Gamma_{\text{r}}/\Gamma_{\text{esc}}$; typically $\Gamma_{\text{esc}}\approx \Gamma_{\text{tr}}$, hence the effective rate is limited by the relaxation process and
becomes temperature dependent. The experimental results from Ref.~\cite{Riwar} indicate that the trapping is relaxation-limited; this implies that making the $S$-$N$ contact more transparent would not affect the trapping rate $\Gamma_\text{eff}$. This holds for a contact in the tunneling regime, so that the proximity effect can be neglected; for good contact between $S$  and $N$, the gap would be suppressed and the trapping would be more appropriately described as being due to gap engineering -- this regime is outside the scope of the present work.

As discussed in the introduction, we are interested in three quantities which are affected by the trap: the relaxation rate of the excess density (i.e., the density which is in addition to the steady-state one), the steady-state density at the junction and its fluctuations.
Focusing on the dynamics of the excess density, we note that the diffusion equation \rref{eq_diff_eq} can be solved in terms of eigenmodes, each with an eigenvalue corresponding to the decay rate of that mode. In general, at long times the slowest mode determines the exponential decay of the excess density, which therefore can be written as
\begin{equation}
x_{\text{qp}}(t, \vec{x})\simeq x_\text{qp}^{0}(\vec{x})e^{-t/\tau_{w}}\ ,
\end{equation}
where the decay rate $\tau_{w}^{-1}$ is the eigenvalue with the smallest absolute value, and $x_\text{qp}^{0}$ its
corresponding eigenfunction. Simple estimates for $\tau_w$ can be found in limiting cases: a weak trap will not significantly affect the density, which can then be taken uniform; after integrating \eref{eq_diff_eq} over the device, we find the decay rate
\be\label{eq_decay_weak}
\frac{1}{\tau_w} \simeq \Gamma_\text{eff} \frac{A_\text{tr}}{A_\text{dev}} \, ,
\ee
where $A_\text{tr}$ and $A_\text{dev}$ are the trap and device area, respectively.
In the opposite regime of a strong trap, the excess density will be fully suppressed at the trap, and its decay rate will be determined by the inverse of the diffusion time between the trap and the region of the device farthest from it; denoting with $L$ the distance of this region from the trap  we have, up to numerical factors of order unity,
\be\label{eq_decay_strong}
\tau_w \simeq L^2/D_\text{qp}\, .
\ee
Let us consider a device with a large aspect ratio, so that $A_\text{dev} = L_\text{dev} w$ with width $w$ much smaller than length $L_\text{dev}$, and a trap of length $d$ and width of the order of $w$. Then assuming a weak trap, using \eref{eq_decay_weak} we find $\tau_{w}^{-1} \sim \Gamma_\text{eff} d/L_\text{dev}$. On the other hand, if the trap is strong and $L_\text{dev}$ is much larger than $d$, \eref{eq_decay_strong} gives $\tau_{w}^{-1} \sim D_\text{qp}/L^2_\text{dev}$. The two rates are comparable when~\cite{Riwar}
\be\label{eq:swco}
d \sim l_0 \equiv \frac{\pi}{2} \frac{\lambda_\text{tr}^2}{L_\text{dev}}\, ,
\ee
with the ``trapping length''
\be\label{lambdatr}
\lambda_\text{tr} = \sqrt{D_\text{qp}/\Gamma_\text{eff}}
\ee
giving the length scale over which the density under the trap decays.
The cross-over from weak ($d\ll l_0$) to strong ($d\gg l_0$) trap was experimentally demonstrated in \ocite{Riwar}.

We point out that the diffusion-limited, strong-trap regime can be reached for traps with dimensions smaller than $\lambda_\text{tr}$ only in the (quasi) one-dimensional geometry. Indeed, let us consider a 2D superconducting film of total area $L_\text{dev}^2$ and a trap of area $d^2$, with $d\ll L_\text{dev}$. In the weak regime the decay rate is $\tau_w^{-1}\approx \Gamma_\text{eff} d^2/L_\text{dev}^2$. Comparing this to the diffusion rate $\sim D_\text{qp}/L_\text{dev}^2$, we find that the crossover from weak to strong trap occurs for the trap size $d\sim\lambda_\text{tr}$. This means that effectively zero-dimensional traps ($d\ll \lambda_\text{tr}$) may be strong in 1D, but they are always weak in 2D. Therefore, it can be advantageous to use quasi-1D devices with small traps, since in 2D devices the traps must be large to be effective, and large traps could potentially lead to unwanted ohmic losses within the normal metal or dissipation at the $S$-$N$ contact. In this paper we focus on the 1D geometry in order to examine the optimal placement of small traps, $d < \lambda_\text{tr}$. Estimation of the trap-induced losses is outside the scope of this work and will be presented elsewhere~\cite{riwarinprep}. However, we note here that the devices measured in \ocite{Riwar} with the longest traps, $d> \lambda_\text{tr}$, had $T_1$ times shorter than those with smaller traps, $d<\lambda_\text{tr}$, giving some evidence for the possible detrimental effect of large traps.

\section{Enhancing the decay rate of the density}
\label{sec:optimal}

In this section we analyze how to optimally place traps of a given size, so that
the slowest mode of the quasiparticle density decays as fast as possible.
As a concrete example, we take the coplanar gap capacitor transmon and
study traps placed in the long wire connecting the gap capacitor to the antenna pads, both via analytical and numerical approaches.
For actual estimates, we use the parameters measured in Refs.~\cite{Riwar,wang}, namely $\Gamma_\text{eff} = 2.42\times10^5$~Hz and $D_\text{qp} = 18$~cm$^2$/s, which, using \eref{lambdatr}, give $\lambda_\text{tr} \simeq 86.2~\mu$m.
Since we are interested in the decay of the excess density, we can set $g=0$; the effect of a trap on the steady-state density due to finite $g$ is the focus of Sec.~\ref{sec:steady}. In Appendix~\ref{app:vortices}, we compare trapping by vortices~\cite{wang} to normal-metal traps.

As we discuss at the end of Sec.~\ref{sec:wiretrap}, considering only the slowest mode for the optimization may not be sufficient when addressing the extreme case of a single, very large trap. However, as we have already pointed out, in a quasi-1D geometry short traps can be strong -- that is, effective at suppressing the excess quasiparticle density. In this case the slowest mode in general still suffices to characterize the long-time quasiparticle decay. The short-trap regime is in particular important for the multiple-trap configurations considered in Sec.~\ref{sec:multitrap}: these configurations combine a fast decay of the quasiparticle density with low electromagnetic losses, and are thus preferable.

\begin{figure}[bt]
 \includegraphics[width=0.48\textwidth]{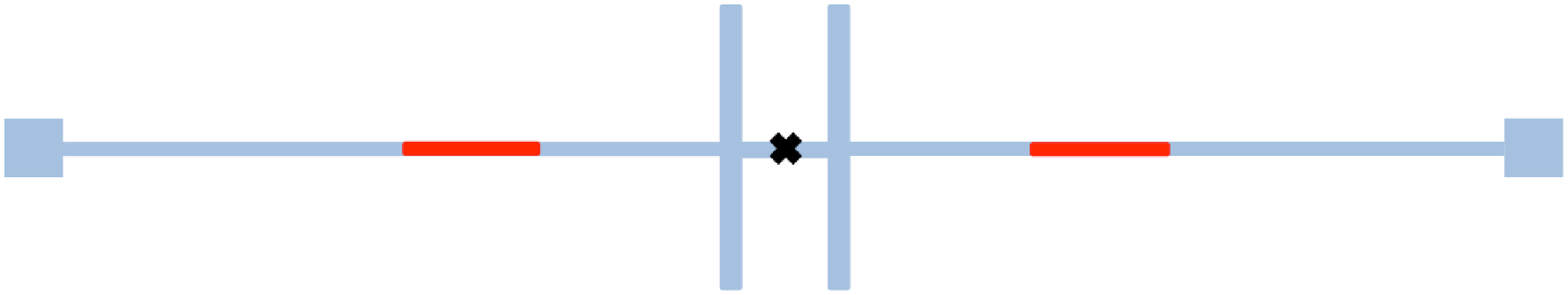}
 \vspace{2mm}
 \includegraphics[width=0.48\textwidth]{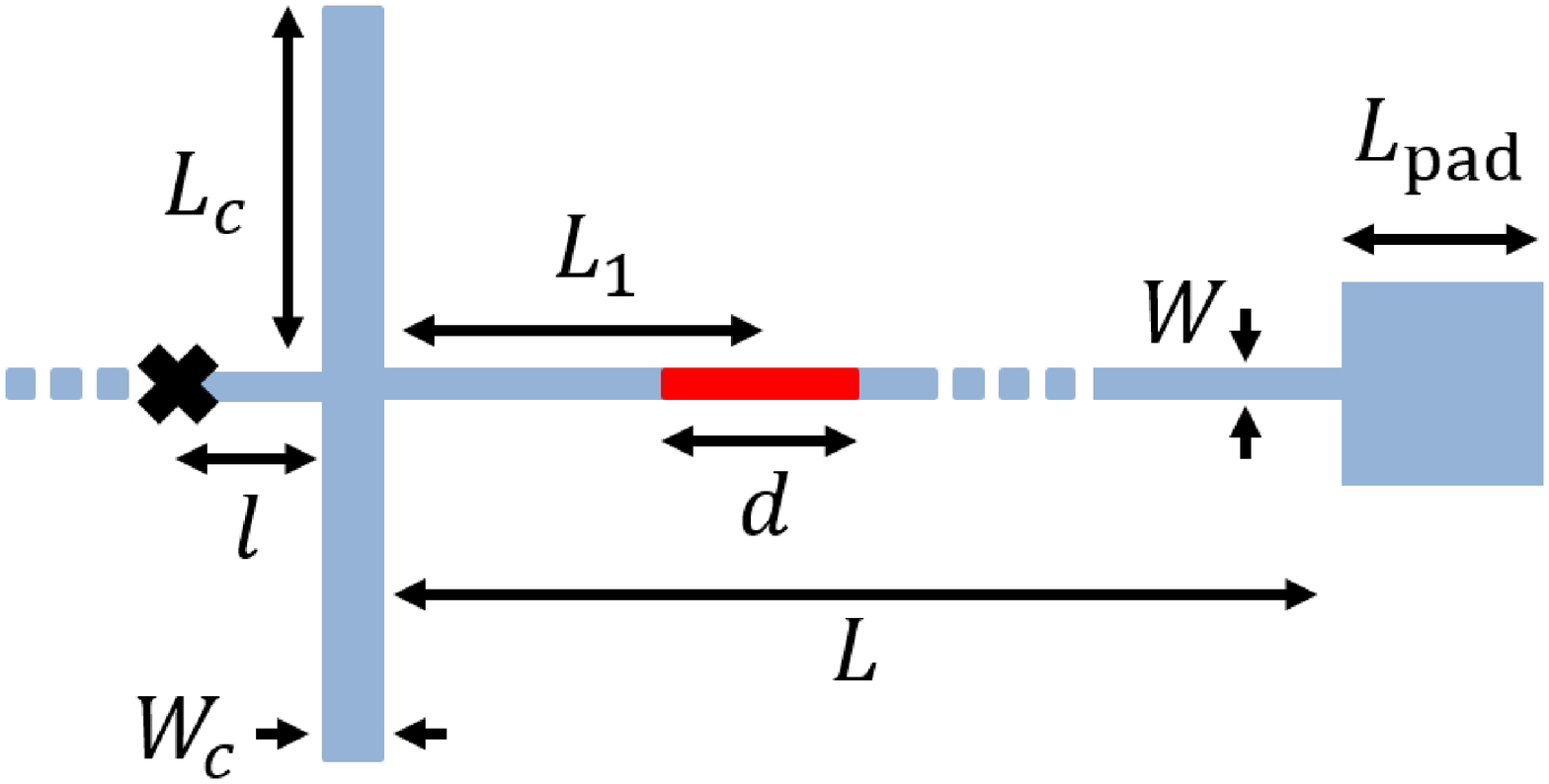}
 \caption{Top: sketch of the transmon qubit studied here, based on the experiments of \ocite{Riwar}. Light blue/light gray: superconducting material; red/dark gray: regions of the superconductor covered by normal metal; cross: position of the Josephson junction. Except for the junction region, the sketch is to scale. Bottom: right half of the device, with the relevant lengths defined: a trap of length $d$ is placed on the antenna wire (length $L$, width $W$) at distance $L_1$ from the gap capacitor (dimensions $L_c$ and $W_c$). The antenna pad is a square of side $L_\text{pad}$.}
 \label{fig:padtrap}
\end{figure}

\subsection{Optimization for a single trap}
\label{sec:wiretrap}

Let us consider a single trap placed in the antenna wire of length $L$, see Fig.~\ref{fig:padtrap} (the device being symmetric, there are two traps in total). We start for simplicity with a short trap of length $d\ll\lambda_\text{tr}$ and neglect the gap capacitor and antenna pads; we then show how to map the full device to this simpler configuration and compare our estimates with numerical results.

For a short trap in a wire, the diffusion equation \rref{eq_diff_eq} can be written in the form (cf. Appendix~\ref{appendix_finite_size_trap})
\begin{equation}\label{eq:short_trap_diff}
\dot{x}_{\text{qp}}=D_{\text{qp}}\vec{\nabla}^{2}x_{\text{qp}}-\gamma_{\text{eff}}\delta\left(y-L_1\right)x_{\text{qp}}\, .
\end{equation}
The trap is at position $y=L_1$ and $\gamma_\text{eff} = d \Gamma_\text{eff}$.
Consider for simplicity the case $\gamma_\text{eff}\rightarrow\infty$, such that quasiparticles are trapped immediately once they reach the trap. As a consequence, $x_\text{qp}(L_1)=0$, and the density on the left and right sides of the trap decays with the rates $\tau_w^{-1}=\pi^2D_\text{qp}/4L_1^2$ and $\tau_w^{-1}=\pi^2D_\text{qp}/4(L-L_1)^2$, respectively. If the trap is at the center of the wire, $L_1=L/2$, the density decays equally fast on both sides, and the decay rate of the slowest mode is four times faster as compared to placing the trap at the beginning or end of the wire. In other words, the central position is the optimal one for the trap to evacuate quasiparticles as quickly as possible.

At finite $\gamma_\text{eff}$, the left/right modes are coupled, but the coupling is small provided that the trap is strong, $d \gg l_0.$ The coupling lifts the mode degeneracy at $L_1=L/2$, but does not change the above conclusion on the optimal position.
We note, however, that if quasiparticles are injected and detected  locally (as, \textit{e.g.}, in \ocite{Riwar}) one may not necessarily observe the global slowest decay rate for strong traps; see Appendix~\ref{app:mode_degeneracy} for more details.

In the simple example above we have shown that the optimal trap position (for which the decay rate of the slowest mode is the fastest) is such that the diffusion times in both sides of the trap are equal. We can extend this finding to a more realistic qubit geometry \cite{Riwar}
which includes the coplanar gap capacitor close to the Josephson junction and the antenna pad
at the far end of the wire, see Fig.~\ref{fig:padtrap}.
The capacitor ``wings'' of length $L_c$ and the square
pad with side $L_\text{pad}$ can be accounted for by adding some effective lengths to the
antenna wire. The effective lengths $L^{\text{eff}}_c\left(k\right)$ and
$L_\text{pad}^{\text{eff}}\left(k\right)$ (for capacitor and pad, respectively) in
general depend on the wave vector $k$, see Appendices \ref{app:Effective-length-pad} and \ref{app:Effective-length-wings}.
If these effective lengths are much smaller than the wire length $L$, we find that for the slow modes the dependence on $k$ drops out: $L^{\text{eff}}_\text{pad}\approx L_{\text{pad}}^{2}/W$ and $L^{\text{eff}}_c\approx 2\frac{W_c}{W}L_{c}$, with $W$ and $W_c$ the widths of the wire and capacitor wings, respectively.
These effective lengths may simply be added to the lengths to the left and right of the trap to find the decay rates:
\be\label{twr}
\frac{1}{\tau_w} = \frac{\pi^2 D_\qp}{4 \left(L-L_1+L_{\text{pad}}^{2}/W - d/2\right)^2}
\ee
for the right mode and
\be\label{twl}
\frac{1}{\tau_w} = \frac{\pi^2 D_\qp}{4 \left(L_1+2\frac{W_c}{W}L_c - d/2\right)^2}
\ee
for the left mode; we have accounted for the finite size of the trap by subtracting the $d/2$ terms in the denominators, and $L_1$ denotes the trap center.
Thus, the optimal trap position is (in the strong trap limit):
\begin{equation}\label{Lopt}
L_{\text{opt}} = \frac{L}{2}+\frac{L_\text{pad}^{\text{eff}}-L_c^{\text{eff}}}{2}\ .
\end{equation}
The optimal position is closer to
the pad (gap capacitor) if the effective length of the pad (capacitor) is larger.

We can check the validity of the above considerations for strong traps and extend our consideration to weaker (i.e., smaller) traps by more accurately modelling the diffusion in the device as done in the Supplementary Information of \ocite{wang} (see also Appendix~\ref{app:vortices}). Namely, the density in the parts not covered by the trap is written in the form
\be\label{xqp_1d}
x_\qp(t,y) = e^{-t/\tau_w} \left[\alpha \cos ky + \beta \sin ky \right] \,
\ee
with $1/\tau_w=D_\qp k^2$
(except for the pad, where the density is assumed uniform), while under the trap we have
\be
x_\qp(t,y) = e^{-t/\tau_w} \left[\alpha \cosh y/\lambda + \beta \sinh y/\lambda \right]\, .
\ee
Imposing continuity of $x_\qp$ and current conservation we find:
\bea
&& z^2 + b^2 = \left(L/\lambda_\text{tr}\right)^2 \, , \label{ptc1}\\
&& \frac{z}{b} [az + \tan \left(z\xi_R\right)]\left[1-h\left(z,\xi_L\right)\tanh\left(b\frac{d}{L}\right)\right]  \qquad \nonumber \\
&& -[1 - az \tan \left(z\xi_R\right)]\left[\tanh\left(b\frac{d}{L}\right)-h\left(z,\xi_L\right)\right]=0, \qquad \, \label{ptc2}
\eea
with $z=kL$, $b=L/\lambda$, $a=L_\text{pad}^2/(LW)$, and
\be\label{f1def}
h\left(z,\xi_L\right)  =\frac{z}{b} \frac{\tan\left(z\xi_L\right)+\tan\left(z\frac{l}{L}\right)+2\frac{W_c}{W}\tan\left(z\frac{L_c}{L}\right)}
{1-\tan\left(z\xi_L\right)\left[\tan\left(z\frac{l}{L}\right)+2\frac{W_c}{W}\tan\left(z\frac{L_c}{L}\right)\right]} .
\ee
We also define the (normalized) length of the wire to the left (right) of the trap by
\bea
\xi_L & = & \left(L_1-d/2\right)/L \, , \\
\xi_R & = & \left(L-L_1-d/2\right)/L \, .
\eea
Solving \esref{ptc1} and \rref{ptc2} for $z$ and $b$, one can find the density decay rate $1/\tau_w = D_\qp z^2/L^2$. For a long qubit with $L\gg \lambda_\text{tr}$, we find for the slow modes $b \approx L/\lambda_\text{tr} \gg 1$ and $z\sim 1$. We note that
the assumption of uniform density in the pad requires $1/\tau_w = D_\qp z^2/L^2 \ll D_\qp/L_\text{pad}^2$; since for experimentally relevant parameters we have $L_\mathrm{pad}^2/L^2 \sim 10^{-2}$, the assumption is valid for slow modes even when $z\gtrsim 1$.

\begin{figure}[bt]
 \includegraphics[width=0.48\textwidth]{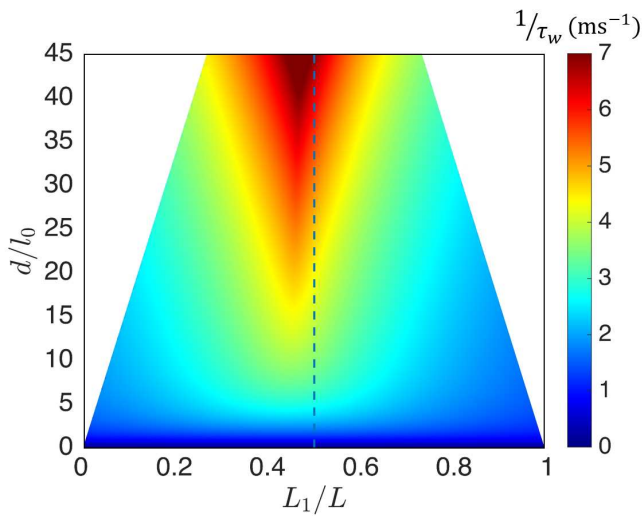}
 \caption{Trapping rate $1/\tau_w$ as a function of the trap position $L_1$ (in units of $L$) and normalized trap size $d/l_0$ -- see Fig.~\ref{fig:padtrap} for the device geometry; the device parameters are: $L=1\,$mm, $l=60\,\mu$m, $W=12\,\mu$m, $L_c=200\,\mu$m, $W_c=20\,\mu$m, $L_\text{pad}= 80\,\mu$m. We used $\lambda_\text{tr} = 86.2\,\mu \mathrm{m}$ for the trapping length, so $l_0= \pi\lambda_\text{tr}^2/2L\simeq 11.7\,\mu$m. The white areas are regions in which the trap center cannot be pushed closer to or further away from the gap capacitor due to the finite trap size.}
 \label{fig:devb1}
\end{figure}

In Fig.~\ref{fig:devb1} we show a density plot of the decay rate $1/\tau_w$ as a function of the distance $L_1$ between gap capacitor and trap center and of the normalized trap size $d/l_0$, calculated using typical experimental parameters as detailed in the caption.
For a strong trap, $d\gg l_0$, as discussed in Sec.~\ref{sec:model} we find that the decay rate is sensitive to the trap position.
The optimum position is shifted with respect to the middle of the wire (dash-dotted line in Fig.~\ref{fig:devb1}), in agreement with the prediction of \eref{Lopt}. Indeed, for the parameters in Fig.~\ref{fig:devb1}, we find $L_\text{pad}^{\text{eff}}=L_{\text{pad}}^{2}/W\approx 533\,\mu\text{m}$ and $L_c^{\text{eff}}=2\frac{W_c}{W}L_{c}\approx 667\,\mu\text{m}$, and the optimal position is closer to the gap capacitor.

\begin{figure}[bt]
 \includegraphics[width=0.48\textwidth]{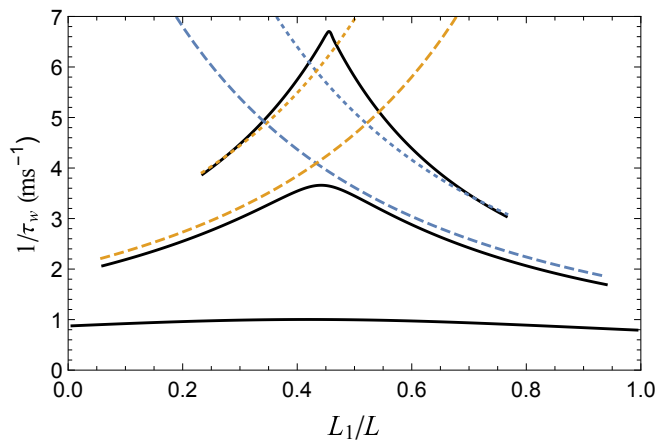}
 \caption{Solid lines: trapping rate $1/\tau_w$ as a function of the trap position $L_1$ measured in units of $L$ for (top to bottom) $d/l_0 = 40,\, 10, \, 1$; other parameters are specified in the caption to Fig.~\ref{fig:devb1} and the device geometry is shown in Fig.~\ref{fig:padtrap}. The dashed lines are the estimates provided by \esref{twr} and \rref{twl} for $d/l_0 = 40,\, 10$.}
 \label{fig:1trna}
\end{figure}

When the trap size is $d\lesssim l_0$, the trap position has only a minor effect on the trapping rate. This is more clearly seen in Fig.~\ref{fig:1trna} (bottom solid curve). For longer traps, we compare the decay from the numerical solution to \esref{ptc1} and \rref{ptc2} with \esref{twr} and \rref{twl}. When the trap is strong but still short compared to the wire (middle solid), \esref{twr} and \rref{twl} (dashed) provide a good approximation to the numerical results (in fact, one can expect the numerically calculated rate to be slower than the analytical prediction, since the numerics account for the finite trapping length which allows for finite density under the trap as well as for the bridge of length $l$ joining the junction to the gap capacitor). For very long traps (upper solid line) the approximation that the effective lengths are small compared to the (uncovered part of the) wire fails, and the calculated rate is faster; this is qualitatively in agreement with the fact that as the mode wavelength increases, the effective lengths
decrease, see \esref{eq:Lepk} and \rref{eq:Leck} (for the gap capacitor, this is true so long as $2W_c/W >1$).

Our focus so far has been in speeding up the decay rate of the slowest mode, without taking into consideration the amplitude of the mode at the junction. This approach is correct for weak traps, $d\lesssim l_0$, since the amplitude of the mode is approximately the same on both sides of the trap. For strong but small traps, $l_0 \ll d \lesssim \lambda_{tr}$, the amplitude on one side of the trap is algebraically suppressed by a factor of order $d/l_0$ compared to the amplitude on the other side (see Appendix~\ref{app:mode_degeneracy}), while for long traps, $d\gg \lambda_\text{tr}$ the suppression is exponential in $d/\lambda_\text{tr}$ (see Appendix~\ref{appendix_finite_size_trap}). In the latter case, it would clearly be advantageous to place the trap close to the junction: the mode with large amplitude between junction and trap would decay quickly, while the slow mode with large amplitude on the other side of the trap would decay slowly but it would be exponentially suppressed at the junction. However, as pointed out at the end of Sec.~\ref{sec:model}, long traps could be too lossy -- this motivates us to further study how to obtain the fastest possible decay using only small traps.

\subsection{Multiple traps}
\label{sec:multitrap}

We now generalize the considerations of the previous section to the case of multiple traps (in each half of the qubit). For a weak trap, the effective trapping rate is proportional to the trap size [\eref{eq_decay_weak}] but independent of position; therefore, no change in the density decay rate can be expected by dividing a weak trap into smaller ones, since the total size is unchanged. The strong-trap regime is qualitatively different in this regard. Let us consider $N_\text{tr}$ strong traps in a wire of length $L$; the traps separate the wire into $N_\text{tr}+1$ compartments. The optimal trap placement is obtained when the diffusion time is the same for each compartment, meaning that the traps have to be placed at positions $L_n=(2n-1)L/2N_\text{tr}$ with $n=1,\ldots,N_\text{tr}$, and
the resulting decay rate is
\begin{equation}\label{eq_multi_trap_decay_rate}
\tau_w^{-1}(N_\text{tr})=N_\text{tr}^2D_\text{qp}\frac{\pi^2}{L^2}\ .
\end{equation}
The rate increases quadratically with the number of traps, so splitting a single strong trap into smaller pieces can highly increase the decay rate. However, when keeping the total area of the traps constant, there is a limitation to this improvement. Indeed, the length of each trap decreases as $d/N_\text{tr}$ and the ``device length'' of each compartment is of order $L/2N_\text{tr}$; using these quantities in \eref{eq:swco} we find that the traps cross over to the weak regime for
\be\label{Nopt}
N_\text{tr}^\text{opt} \sim \sqrt{\frac{d}{2l_0}} \, ;
\ee
here $l_0$ is defined by the right hand side of \eref{eq:swco} with $L_\text{dev} = L$. Increasing the trap number beyond $N_\text{tr}^\text{opt}$ does not further improve the decay rate, which is thus limited by \eref{eq_decay_weak}. In other words, to obtain that maximum decay rate for given total length $d$, at least $N_\text{tr}^\text{opt}$ traps should be placed evenly spaced over the device.
Such a configuration could also reduce the trap-induced losses, since they depend on the trap position and size~\cite{riwarinprep}.

Let us now show in a concrete example that multiple traps can indeed increase the decay rate as predicted by \eref{eq_multi_trap_decay_rate}.
We consider again the transmon device depicted in Fig.~\ref{fig:padtrap}, but
we now assume that two traps are placed on the central wire, with distances $L_1$ and $L_2$ between the gap capacitor and the traps centers.
Accounting for the second trap, we generalize \eref{ptc2} to
\be \label{2tde} \begin{split}
& \frac{z}{b}\left[1-h(z,\xi_L)\tanh\left(b\frac{d_1}{L}\right)\right]\bigg\{
 \frac{z}{b}\tan \left(z\chi\right) -\tanh\left(b\frac{d_2}{L}\right) \\& + g(z,\xi_R)\left[1-\frac{z}{b}\tan \left(z\chi\right)\tanh\left(b\frac{d_2}{L}\right)\right]\bigg\} \\& -\left[\tanh\left(b\frac{d_1}{L}\right)-h(z,\xi_L)\right]
\bigg\{\tanh\left(b\frac{d_2}{L}\right)\tan \left(z\chi\right) \\ & + \frac{z}{b} -g(z,\xi_R)\left[\tan \left(z\chi\right) + \frac{z}{b}\tanh\left(b\frac{d_2}{L}\right)\right]\bigg\} = 0
\end{split}\ee
where, taking $L_1 < L_2$,
\bea
\xi_L & = & (L_1 - d_1/2)/L\, , \\
\xi_R & = & (L - L_2 -d_2/2)/L\, , \\
\chi & = & (L_2-d_2/2 - L_1 - d_1/2)/L \, ,
\eea
the function $h(z,\xi_L)$ is defined in \eref{f1def}, and
\be
g(z,\xi_R) = \frac{z}{b} \frac{az + \tan (z\xi_R)}{1-az\tan (z\xi_R)}\ .
\ee

\begin{figure}[bt]
 \includegraphics[width=0.4\textwidth]{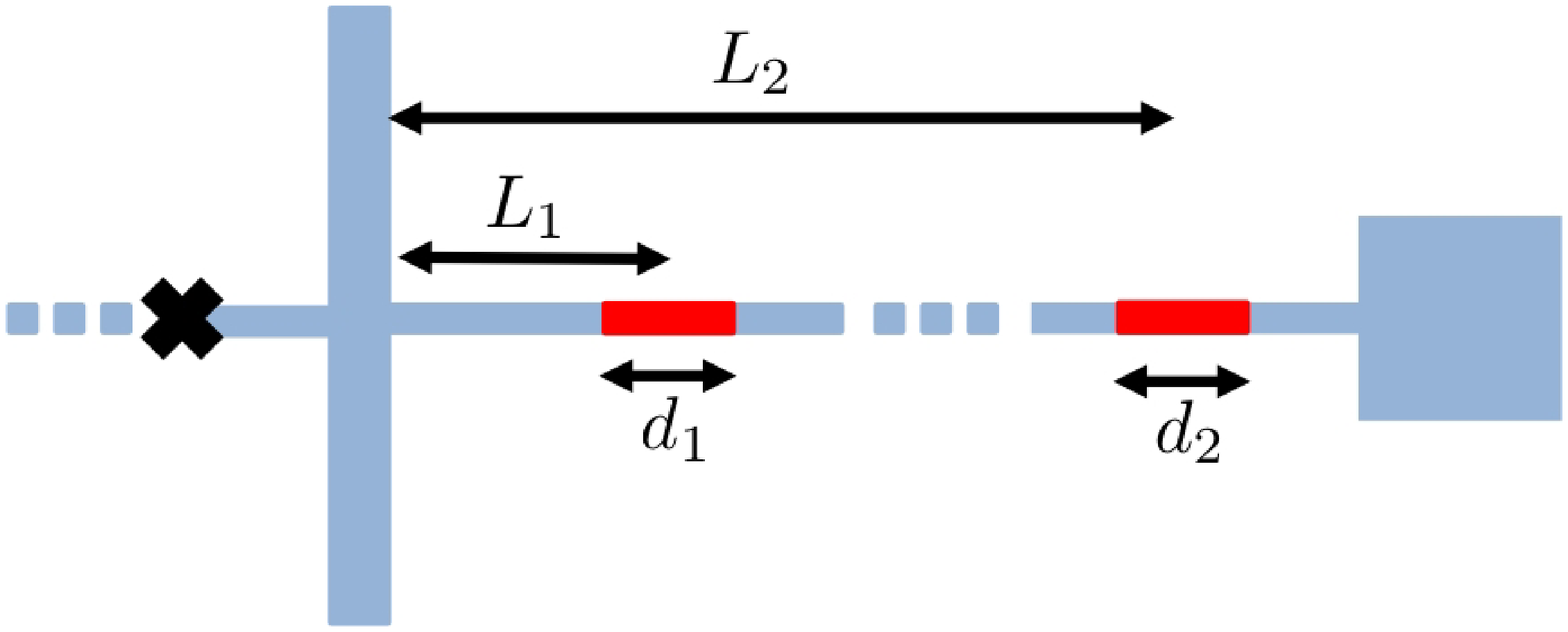}
\vspace{-7mm}\flushleft{(a)}
 \includegraphics[width=0.48\textwidth]{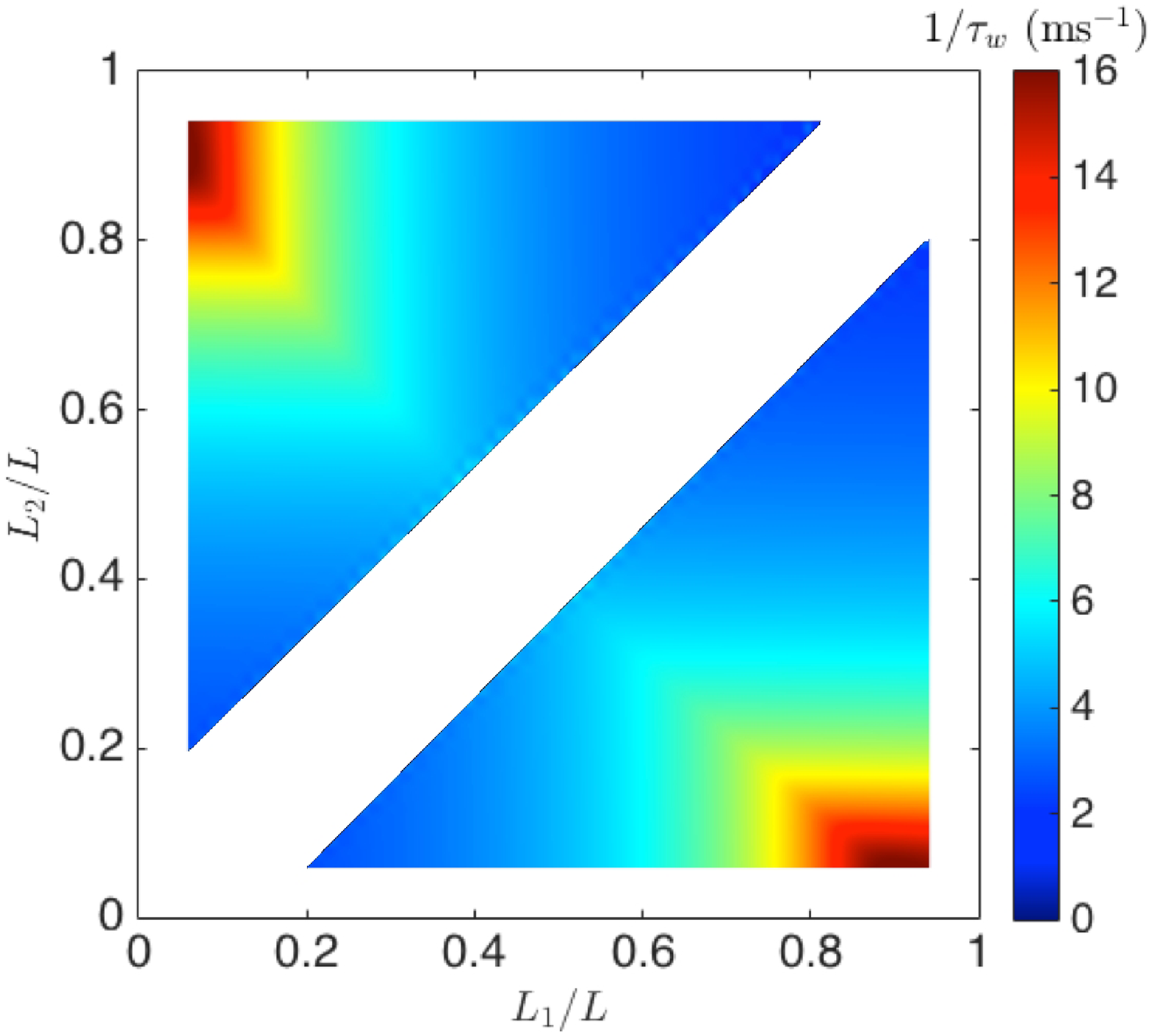}
\vspace{-11mm}\flushleft{(b)}
 \caption{(a) Device with two traps (dark red) in each half of the qubit; distances $L_1$ and $L_2$ are measured from the gap capacitor to the center of each trap, cf. Fig.~\ref{fig:padtrap}. (b) Trapping rate $1/\tau_w$ as function $L_1$ and $L_2$; here the two traps are identical, $d_1=d_2=10l_0$, which makes the plot symmetric under the  exchange $L_1 \leftrightarrow L_2$. The parameters used are specified in the caption to Fig.~\ref{fig:devb1}. Similar to that figure, the white areas correspond to forbidden regions due to the finite traps' sizes . A comparison with Fig.~\ref{fig:devb1} reveals that splitting a trap with length $d=20l_0$ into two identical ones can boost the trapping rate up to a factor larger than 3.}
 \label{fig:dp}
\end{figure}

We consider for simplicity the case of equal traps, $d_1=d_2\equiv d/2$ with $d=20 l_0\simeq 233\,\mu$m the total length of the normal metal.
We show in Fig.~\ref{fig:dp} the decay rate as function of $L_1$ and $L_2$  for the same parameters as in Fig.~\ref{fig:devb1}.
We find that the decay rate of the slowest mode is highest when placing the traps far away from each other, one trap touching the gap capacitor and the other begin close to the pad.
This is in qualitative agreement with our expectations: consider again the gap capacitor and the pad as extra lengths added to the left and right of the central wire, respectively; this leads to a wire of effective total length $L_\text{tot}=L^\text{eff}_c+ L+L^\text{eff}_\text{pad}\simeq 2200\,\mu$m. In such a wire the optimal positions would be $L_1 = L_\text{tot}/4 \simeq 550\,\mu$m and $L_2 = 3L_\text{tot}/4 \simeq 1650\,\mu$m. The value of $L_1$ would indicate an optimal position inside the gap capacitor, but since we allow for the traps to be placed in the central wire only, this optimal placement is not possible. The value of $L_2$ corresponds to a position slightly away from the pad, in agreement with the results in Fig.~\ref{fig:dp}. Finally, going from the optimally-placed single trap to the optimal two-trap configuration, the decay rate increases by a factor of $\sim 3.4$. This factor does not reach the theoretical maximum of 4 predicted by \eref{eq_multi_trap_decay_rate}; the discrepancy can be attributed both to the non-optimal placement of the first trap mentioned above as well as to finite-size effects, as in the single trap case. However, the calculated improvement confirms that the decay rate can be significantly increased by optimizing the trap number and position.

It is instructive to compare our results for the two-trap case with the fast decay of the mode to the left of the single trap; using \eref{twl}, we estimate the decay rate of this mode to be $1/\tau_w \simeq 14.7\,$ms$^{-1}$, slightly slower than the maximum rate shown in Fig.~\ref{fig:dp}. In both cases, we do not allow the trap to enter the gap capacitor; this constraint could be important in limiting trap-related losses, as the gap capacitor is the region with the highest electric field. Using two traps we place only half the total normal material near the high-field region, while obtaining a slightly faster decay than with a single, large trap.
For the considered example, further increase in the decay rate could be obtained by further splitting the traps. Indeed, a more accurate estimate for the length $l_0$ can be obtained by using $L_\text{dev} = L_\text{tot} - d \simeq 1967\,\mu$m in \eref{eq:swco}, giving $l_0\approx 5.9\,\mu$m. Then the ``optimal'' trap number would be $N_\text{tr}^\text{opt} \simeq \sqrt{d/2l_0} \sim 4$, requiring one trap to be placed on the pad, two on the antenna wire, and one on the gap capacitor (this placement is calculated using the ``effective wire'' length of the gap capacitor, so that in practice one should symmetrically place one trap on each of the two ``wings'' of gap capacitor). As mentioned above, placement on the gap capacitor could be detrimental, but with the optimal trap number only one quarter of the normal metal would be in the gap capacitor and the resulting losses would therefore be smaller than those due to a single large trap on the gap capacitor. Therefore, it is potentially beneficial to have multiple smaller traps in comparison with a single large trap. Such considerations are also dependent on the device design. 
In Appendix~\ref{app:Xmon} we briefly consider a different geometry for the qubit, the Xmon of Ref.~\onlinecite{barends}; the central X-shaped part of the device is small, and unfortunately this implies that no large gain in the decay rate can be obtained using multiple traps, so alternative approaches are desirable in this case.
In the next section we turn our attention to the effect of traps on the steady-state density.

\section{Suppression of steady-state density and its fluctuations}
\label{sec:steady}

In the preceding section we have dealt with the question of how fast quasiparticles reach their steady state if there is a deviation from said steady-state density. In this section we point out that traps also affect the shape of the steady-state density. In particular, our aims are to minimize the steady-state density at the junction, which directly affects the $T_1$ time of qubits, as well as to stabilize the density value against fluctuations in their generation rate which lead to temporal variations in the qubit lifetime.
In our model, the steady-state density is nonzero due to a finite generation rate $g$ in Eq.~\eqref{eq_diff_eq}. As argued in Sec.~\ref{sec:model}, in the presence of a weak trap
the quasiparticle density is uniform, and in the steady-state takes the value $x_\qp^s = g \tau_w$ with $\tau_w$ of \eref{eq_decay_weak}.
As we now show, going beyond the weak limit the geometry affects the spatial profile of the density.

For a concrete example, we consider the same geometry as in Sec.~\ref{sec:wiretrap} -- that is, a single trap on the wire connecting gap capacitor and pad, see Fig.~\ref{fig:padtrap}. The solution for the profile of the steady-state density in each 1D segment outside the trap is given by parabolas of the general form $x_\qp^s = -y^2 g/2D_\qp +\alpha y + \beta$, while under the trap we have $x_\qp^s = \tilde\alpha \cosh(y/\lambda_\text{tr})+ \tilde\beta \sinh (y/\lambda_\text{tr}) + g/\Gamma_\text{eff}$. The pad density is again assumed constant. The parameters $\alpha$, $\beta$ in each segment, as well as $\tilde\alpha$, $\tilde\beta$ are found by imposing appropriate boundary conditions (i.e., continuity and current conservation).
We finally arrive at the following expression for the steady-state density $x_\qp^J$ at the junction:
\be\label{xqpj2}\begin{split}
x_\qp^J  = &\frac{g}{\Gamma_{\mathrm{eff}}}
\left[1+\frac{1}{\sinh (d/\lambda_\text{tr})}\frac{A_R}{W \lambda_\text{tr}} + \coth \left(d/\lambda_\text{tr}\right)\frac{A_L}{W\lambda_\text{tr}} \right]
\\ & + \frac{g}{D_\qp} \left[ \frac{(L_1+l-d/2)^2}{2} +\frac{A_c(L_1-d/2)}{W} \right],
\end{split}\ee
where $A_R=W[L-L_1-d/2]+L_\text{pad}^2$ and $A_L=W[L_1+l-d/2]+A_c$ are the uncovered areas to the right and left of the trap, respectively, and $A_c=2W_cL_c$ is the gap capacitor area.

In the small trap limit, $d \ll \lambda_\text{tr}$,  we can rewrite \eref{xqpj2} in the form
\be\label{xqpj3}
x_\qp^J \simeq g \left( \tau_w + t_D \right)
\ee
with $\tau_w$ defined in \eref{eq_decay_weak}, while
$t_D = [(L_1+l-d/2)^2/2 + A_c(L_1-d/2)/W ]/D_\qp$ represents the diffusion time between junction and trap (with the second term in square brackets taking into account the presence of the gap capacitor). Similar to the discussion in Sec.~\ref{sec:model}, we can distinguish between an effectively weak ($\tau_w \gg t_D$) and strong trap ($\tau_w\ll t_D$), with the trap becoming strong as its length $d$ increases above the position-dependent length scale $l_1 \sim\lambda_\text{tr}^2/\sqrt{D_\qp t_D}$. Note that $l_1$ decreases with the distance $L_1$ between gap capacitor and trap and is always larger than $l_0$ of \eref{eq:swco}; therefore a trap that is weak in the sense of $d$ being smaller than $l_0$ is weak at any position $L_1$, and the value of $x_\qp^J$ is only weakly dependent on the trap placement. On the other hand, a strong trap with $d>l_0$ effectively becomes weak, as $L_1$ decreases, when
$\tau_w = t_D$.
At positions $L_1$ smaller than that given by this condition, $x_\qp^J$ again becomes weakly dependent on trap placement. In other words, the condition determines the maximal distance at which the largest (up to numerical factor) suppression of $x_\qp^J$ is achieved for a given trap size $d$.

\begin{figure}[bt]
 \includegraphics[width=0.48\textwidth]{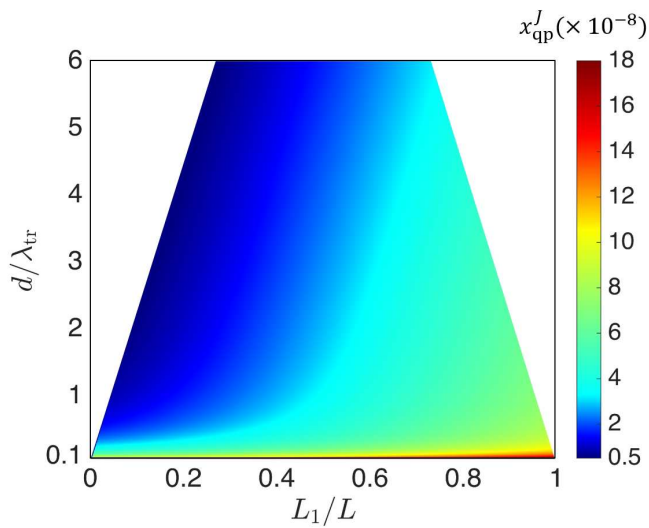}
 \caption{Quasiparticle density at the junction as a function of trap location $L_1$ (in units of $L$) and its normalized size $d/\lambda_\text{tr}$, calculated using $g=10^{-4}\,$Hz~\cite{wang,vool}; the normalization of the size $d$ differs from that of Fig.~\ref{fig:devb1}, but the parameters used for the device are the same specified there.}
 \label{fig:deatja}
\end{figure}

For a long trap $d\gg \lambda_\text{tr}$, we can still use \eref{xqpj3} after the identification
\be\label{longtraptau}
\tau_w \to \frac{1}{\Gamma_\text{eff}} \left(1 + \frac{A_L}{W \lambda_\text{tr}}\right)\, .
\ee
With this substitution and for typical experimental parameters, we find again that the first term in \eref{xqpj3} dominates when the trap is close to the junction (despite being smaller than the corresponding term for a short junction), while the second one takes over as $L_1$ increases. More generally, it should be noted that in any regime $x_\qp^J$ is a monotonically increasing function of $L_1$: as one could expect, the closer the trap is to the junction, the more it suppresses the quasiparticle density near the latter.
This behavior is evident in Fig.~\ref{fig:deatja}: the plot clearly shows that the density is suppressed by placing the trap near to the gap capacitor, and that long traps ($d\gtrsim \lambda_\text{tr}$) are more effective. Values as low as $x_\qp \sim 10^{-8}$ are predicted; for comparison, we note that in devices with the geometry considered here but without traps, we estimate $x_\qp \sim 10^{-6}$~\cite{wang}. On the other hand, for transmons with larger pads (so that there are always vortices that act as traps) we find $x_\qp \sim 10^{-7}$~\cite{3dtr,wang}. Further suppression of the density could be achieved by placing traps in the gap capacitor, since this would effectively reduce the uncovered area $A_L$ [cf. \eref{longtraptau}] between trap and junction.

Based on the above consideration, we do not expect that adding a second trap far from the junction significantly affects the steady-state density $x_\qp^J$ (we have confirmed this by direct calculation). Therefore, the results of the last two sections suggest that having two traps, one close to the junction and the other close to the pad can both greatly reduce the steady-state density at the junction and enhance the decay rate of the excess density. Next, we show that traps can also contribute to the temporal stability of
the qubit.

\subsection{Fluctuations in the generation rate}
\label{sec:fluctuations}

As discussed previously, the density at the junction and hence the qubit relaxation rate are proportional to the generation rate $g$. Therefore, temporal variations in $g$ can cause changes in the measured $T_1$ over time. Here we explore how traps can suppress these changes. For this purpose,
we include Gaussian fluctuations of $g$ in \eref{eq_diff_eq} by replacing
$g\rightarrow g+\delta\widehat{g}\left(y,t\right)$, with
\begin{align}
\left\langle \delta\widehat{g}\left(y,t\right)\right\rangle  & = 0 \, ,\\
\left\langle \delta\widehat{g}\left(y,t\right)\delta\widehat{g}\left(y',t'\right)\right\rangle  & = \gamma_{g}\delta\left(y-y'\right)\frac{1}{2\tau_{m}}e^{-\frac{\left|t-t'\right|}{\tau_{m}}}.
\end{align}
The parameter $\gamma_{g}$ characterizes the fluctuation amplitude
and has the same units as $\gamma_{\text{eff}}$ of \eref{eq:short_trap_diff}, while $\left\langle \ldots\right\rangle $
averages over all realizations of $\delta\widehat{g}$. We assume that fluctuations are spatially uncorrelated, but allow for temporal correlations with a finite memory time $\tau_{m}$ -- we will return to this point in what follows. Note that under these assumptions the average density $\langle x_\qp (y) \rangle$ is in general a function of the spatial coordinate due to the presence of traps, but not of time.

To provide a measure for the density fluctuations at the junction position $y=0$, we consider
the quantity
\begin{equation}
\Delta x_{\text{qp}}^2\left(t,t'\right)\equiv\left\langle x_{\text{qp}}\left(0,t\right)x_{\text{qp}}\left(0,t'\right)\right\rangle -\left\langle x_{\text{qp}}\left(0\right)\right\rangle^2  .\label{eq:definition_Delta_x_qp}
\end{equation}
This quantity can be expressed in terms of the eigenvalues $\mu_k < 0$ and eigenfunctions $n_k(y)$ of \eref{eq_diff_eq} (cf. \ocite{Riwar} and Appendix~\ref{appendix_finite_size_trap}).
Indeed, the quasiparticle density with the fluctuation term is
\begin{equation}
\begin{split}
x_{\text{qp}}\left(y,t\right)= & -\sum_{k}\frac{1}{\mu_{k}}n_{k}\left(y\right)g_{k}\\
& + \sum_{k}\int_{-\infty}^{t}dt_{1}e^{\mu_{k}\left(t-t_{1}\right)}n_{k}\left(y\right)\delta\widehat{g}_{k}\left(t_{1}\right),
\end{split}
\end{equation}
with
\begin{equation}
\delta\widehat{g}_{k}\left(t_{1}\right)=\int_{0}^{L}\frac{dy'}{L}n_{k}\left(y'\right)\delta\widehat{g}\left(y',t_{1}\right),
\end{equation}
and the similar definition for $g_k$. Substituting this expression into the definition \eref{eq:definition_Delta_x_qp}
and averaging over the fluctuations, we find
\begin{equation}
\begin{split}
\Delta x_{\text{qp}}^2\left(t,t'\right)=\frac{\gamma_{g}}{2L}\frac{1}{\tau_{m}}\sum_{k}\int_{-\infty}^{t}dt_{1}\int_{-\infty}^{t'}dt_{2}\, e^{\mu_{k}\left(t-t_{1}\right)}\\ \times e^{\mu_{k}\left(t'-t_{2}\right)}e^{-\frac{\left|t_{1}-t_{2}\right|}{\tau_{m}}}n_{k}^{2}\left(0\right)\, .
\end{split}
\end{equation}
The time integrals are conveniently computed by first shifting the
times $t_{1}$ and $t_{2}$ by $t$ and $t'$, respectively, and then
changing variables in the two-dimensional integral into a mean time $t_{1}+t_{2}$
and time difference $t_{1}-t_{2}$. After integration, we obtain
\begin{equation}\label{eq:Dxqp2}
\Delta x_{\text{qp}}^2\left(t-t'\right) = \frac{\gamma_{g}}{2L}\sum_{k}\frac{\tau_{m}e^{-\frac{\left|t-t'\right|}{\tau_{m}}}+\frac{1}{\mu_{k}}e^{\mu_{k}\left|t-t'\right|}}{\tau_{m}^{2}\mu_{k}^{2}-1}n_{k}^{2}\left(0\right),
\end{equation}
which depends only on time difference.

Equation \rref{eq:Dxqp2} shows that even in the absence of time correlations for the fluctuations in the generation rate, $\tau_m =0$, the fluctuation in the density are correlated due to diffusion. In this case, the longest decay time is that of the slowest mode, $1/\mu_0$, and is typically of the order of milliseconds~\cite{Riwar}. This time is shorter than the time it takes to measure a qubit relaxation curve and hence estimate the quasiparticle density. Therefore, only the regime in which $\tau_m$ is much longer than $1/\mu_0$ could have observable consequences. Moreover, there is experimental evidence for slow fluctuations in the number of quasiparticles in qubits, obtained by monitoring the quantum jumps between states of a fluxonium, \ocite{vool}, and by repeated measurements, over several hours, of the relaxation time in a capacitively shunted flux qubit, \ocite{gustavsson}. Therefore, in the reminder of this section we focus only on the regime of long memory -- that is, slow fluctuation in the generation rate. In the limit $\tau_m \gg 1/\mu_0$, \eref{eq:Dxqp2} simplifies to
\begin{equation}\label{eq:Dxqp2tm}
\Delta x_{\text{qp}}^2\left(t-t'\right) \simeq \frac{\gamma_{g}}{2L\tau_m}e^{-\frac{\left|t-t'\right|}{\tau_{m}}} \sum_{k}\frac{1}{\mu_{k}^{2}}n_{k}^{2}\left(0\right).
\end{equation}

We now want to establish that a trap can indeed reduce the fluctuations. To this end, we consider the simple case of the junction in a quasi-1D wire extending for length $L$ from the junction and with a trap at distance $L_1$ from the junction. Initially, we take the trap to be small (length $d\ll \lambda_\text{tr}$), and we distinguish between weak and strong trap, see \eref{eq:swco}.
For a weak trap, $d \ll l_0$, the slow modes are only weakly dependent on the spatial coordinate. Moreover, for the slowest mode the decay rate is [cf. \eref{eq_decay_weak}]
\begin{equation}
\mu_{0}\approx -\Gamma_{\text{eff}}\frac{d}{L}\, ,
\end{equation}
while the higher modes are much faster, since $\mu_{n>0}\lesssim -D_{\text{qp}}/L^{2}$,
and thus $\left|\mu_{n>0}\right|\gg\left|\mu_{0}\right|$.
Using $n_{0}(0) \approx 1$ and neglecting the small contributions from the higher modes, from \eref{eq:Dxqp2tm} we find
\begin{equation}\label{Dxqp2w}
\Delta x_{\text{qp}}^2\left(t-t'\right) \approx \frac{\gamma_{g}}{2L\tau_m}e^{-\frac{\left|t-t'\right|}{\tau_{m}}}
\left(\frac{L}{\Gamma_\text{eff}d}\right)^2.
\end{equation}
This expression shows that a stronger/longer trap more effectively suppresses fluctuation, as could be expected.

In the case of a strong trap ($d \gg l_0$), the eigenmodes can be split in two sets (cf. Sec.~\ref{sec:wiretrap}): there are left and right modes, which are strongly suppressed
to the right and to the left of the trap, respectively (here we assume that the trap position is sufficiently far from the central position, see Appendix~\ref{app:mode_degeneracy}). The left modes, with large amplitude between junction and trap, give small contributions to the density fluctuations when the trap is close to the junction and their contributions grow with junction-trap distance. The right modes, while being suppressed to the left of the trap, have opposite behavior with distance, so they can dominate when the trap is sufficiently close to the junction. Then we generically expect a non-monotonic dependence of the density fluctuations on trap-junction distance from the competition between modes to the left and right of the trap.

Let us consider the decay rates for left and right modes, which we denote with $\mu_{n,L_1}$ and $\mu_{n,L-L_1}$, respectively, where we define
\begin{equation}
\mu_{n,\ell}\simeq -D_\qp \left(\frac{\pi}{2\ell}\right)^{2}\left(2n+1\right)^{2}, \ n=0,\,1,\,\ldots
\end{equation}
Keeping in \eref{eq:Dxqp2tm} only the slowest mode for each set, since
the higher modes with $n>0$ gives a smaller contribution to the sum, we find
\be\label{Dxqp2s}\begin{split}
\Delta x_{\text{qp}}^2\left(t-t'\right) \approx \frac{\gamma_{g}}{L\tau_m}e^{-\frac{\left|t-t'\right|}{\tau_{m}}}\frac{1}{\mu_{0,L}^2}
 \\ \times\left[\left(\frac{L_1}{L}\right)^3 + \left(\frac{l_0}{d}\right)^2\frac{L-L_1}{L}\right]\, ,
\end{split}\ee
where we used $n_{k,L_1}^2(0) \simeq 2L/L_1$ and $n_{k,L-L_1}^2(0) \simeq 2 (l_0/d)^2[L/(L-L_1)]^3$ (here we also assume $d\ll L_1,\, L-L_1$).
The first term in square brackets originates from the lowest mode confined between junction and trap, while the second term, due to the lowest mode located on the other side of the trap, is suppressed by the small factor $(l_0/d)^2$.
As a function of the trap position $L_1$, in agreement with the above considerations we find that $\Delta x_{\text{qp}}^2$ has a minimum at $L_1 = L l_0/d\sqrt{3}$, where the terms
in square brackets take the approximate value $(l_0/d)^2$ and \eref{Dxqp2s} takes the same form of \eref{Dxqp2w}. In fact, those terms rise significantly above this value only for $L_1 > L_f \equiv L (l_0/d)^{2/3}$, indicating that for a strong trap a large suppression of fluctuations can be achieved if the trap is not placed far beyond $L_f$. We note that this condition is more stringent than the one discussed after \eref{xqpj3}, $\tau_w = t_D$, which for the simple wire considered here gives a maximum distance $\sim 2L\sqrt{l_0/\pi d}$; in other words, maximum suppression of fluctuations ensures maximum suppression of the steady-state density.

The above considerations for a strong but short trap can be generalized to longer traps ($d \gtrsim \lambda_\text{tr}$, with $d\ll L$) by substituting $l_0/d \to \lambda_\text{tr}/L\sinh(d/\lambda_\text{tr})$, see Appendix~\ref{appendix_finite_size_trap}. In both regimes (strong but short, and long trap), increasing the trap length suppresses the fluctuations at the junction, but shrinks the region over which maximum suppression can be achieved, since $L_f$ becomes smaller. This region is however always small compared to the wire length, $L_f \ll L$. Together with the monotonic dependence of the average quasiparticle density on distance obtained in the first part of this section, our results show that placing a trap close to the junction is effective in suppressing both the average density and its fluctuations, potentially making the qubit longer lived and more stable.

\section{Summary}
\label{sec:summary}

In this paper we study the effects of size and position of normal-metal quasiparticle traps in superconducting qubits with large aspect ratio, so that quasiparticle diffusion can be considered one-dimensional. We focus on such a design because, as we argue at the end of Sec.~\ref{sec:model}, in a two-dimensional setting traps must be large compared to the trapping length $\lambda_\text{tr}$ [\eref{lambdatr}] to be strong, while in quasi-1D it is sufficient for the trap length $d$ to be longer than the characteristic scale $l_0$ [\eref{eq:swco}] which accounts for diffusion, trapping rate, and device size -- this characteristic scale is generally shorter than $\lambda_\text{tr}$ for long devices ($L_\text{dev}>\lambda_\text{tr}$). A trap can influence the qubit in three ways: first, it suppresses the steady-state quasiparticle density at the junction; then the  qubit's $T_1$ time can be increased, since this time is inversely proportional to the density. Second, a trap can speed up the decay of the excess quasiparticles and, third, it can decrease fluctuations around the steady-state density; these effects can render the qubit more stable in time -- in fact, there is experimental evidence (Refs.~\cite{vool} and \cite{gustavsson}) that fluctuations in the number of quasiparticles are responsible for at least part of the temporal variations in $T_1$.
Not surprisingly, a long trap ($d \gtrsim \lambda_\text{tr}$) placed close to the junction is effective in all three aspects: fast decay of excess quasiparticles, suppression of the steady-state quasiparticle density (see Fig.~\ref{fig:deatja}), and suppression of density fluctuations at the junction. However, large traps could be a source of unwanted dissipation; therefore, we analyze in more detail the effects of shorter traps.

If a trap is weak, $d\lesssim l_0$, its position has little influence on the ability to suppress the quasiparticle density and its fluctuations, as well as on the decay rate of excess quasiparticles. Interestingly, for a strong but short trap, $l_0\lesssim d \lesssim \lambda_\text{tr}$, we find the position of the trap can be optimized in several ways. First, there is an optimal position that makes the decay of excess quasiparticles as fast as possible, see Figs.~\ref{fig:devb1} and \ref{fig:1trna} in Sec.~\ref{sec:wiretrap}; however, a better choice is in general to divide a strong trap into smaller traps of length $\sim l_0$ and distribute those around the device, see Sec.~\ref{sec:multitrap}. For the suppression of density fluctuations, we find that there is an optimal trap position, see Sec.~\ref{sec:fluctuations}; more importantly, we find that there is a maximum distance $L_f$ from the junction up to which the suppression of fluctuations is effective. Moreover, the distance up to which a large suppression of the steady-state density is achieved (Sec.~\ref{sec:steady}) is longer than $L_f$, so that suppressing fluctuations also suppresses the steady-state density. Therefore, by correctly placing multiple traps in the device in such a way that one is sufficiently close to the junction, all three beneficial effects of traps can be optimized.


The optimization of trap size, number, and placement is the only readily accessible way to improve the trap efficacy, since the effective trapping rate is limited by the energy relaxation rate in the normal metal~\cite{Riwar}, a material parameter that cannot be easily modified. We stress here that these considerations are valid for normal islands in tunnel contact with the superconductor -- traps formed by gap engineering (\textit{e.g.}, by placing a lower-gap superconductor in good contact with the qubit) could behave differently and deserve further consideration.


\acknowledgments

We gratefully acknowledge interesting discussions with L. Burkhart, Y. Gao, I. Khaymovich, and A. Rezakhani.
This work was supported in part by the EU under REA Grant Agreement No. CIG-618258 (G.C.), ARO Grant W911NF-14-1-0011 and a Max Planck award (R.P.R., R.J.S.), and DOE contract DEFG02-08ER46482 (L.G.).

\appendix

\section{Comparison with vortex trapping}
\label{app:vortices}

In the coplanar gap capacitor transmon, the antenna pads are the widest part of the device. This makes it possible to trap vortices only in the pads when cooling the device in a small magnetic field. It was shown in \ocite{wang} that each vortex added to a pad increases the density decay rate, and the effectiveness of trapping by vortices was characterized by a ``trapping power'' $P$. Here we compare the vortex trapping with a normal-metal trap covering the pad.

To determine the decay rate of the excess density $x_\qp$, we construct the solution to the diffusion equation \rref{eq_diff_eq}, along the lines of the Supplementary to \ocite{wang}. We treat all parts of the device except the pad as one-dimensional and write $x_\qp$ in each segment in the form of \eref{xqp_1d}.
We approximate the density in the pad as uniform (justified for the lowest mode if $L_\text{pad}<\lambda_\text{tr}$, as in the actual devices).
We then impose continuity of the density and current conservation where the parts of the device meet and thus arrive at
the following effective boundary condition for the density at the connection between wire and  pad (at $y=L$):
\be\label{eqxbou}
\frac{\partial x_\qp}{\partial y}\Big|_{y=L}=L_{\mathrm{pad}}^2 \bigg[ \frac{D_\qp k^2 - \Gamma_{\mathrm{eff}}}{WD_\qp}\bigg] x_\qp(y=L)\ .
\ee
Let us introduce for simplicity the dimensionless parameter $z=kL$; after imposition of all boundary conditions, as detailed in \ocite{wang}, we find that the parameter must satisfy the equation
\be\label{eqzpadtrap}
\bigg ( \frac{L_{\mathrm{pad}}^2\Gamma_{\mathrm{eff}}t_L}{LW}  \bigg ) [1-f(z)\tan z]-z[\tan z +f(z)]=0\ ,
\ee
with $t_L = L^2/D_\text{qp}$ and (see Ref.~\cite{wang})
\be
f(z)=\tan \left(z\frac{l}{L}\right) + 2 \frac{W_c}{W} \tan\left( z \frac{L_c}{L}\right)\ .
\ee

The similar calculation for the case of $\bar{N}$ vortices in each pad leads to the following equation for $z$:
\be\label{zeqv}
\left(\bar{N} \frac{P t_L}{LW} -a z^2\right) [1-f(z)\tan z]-z[\tan z +f(z)]=0
\ee
with $a=L_\mathrm{pad}^2/(LW)$; since
in the experiments $a$ as well as $z$ for the lowest mode and the coefficient multiplying $\bar{N}$ are all of order unity, in this equation we can neglect the term proportional to $a$  for large number of vortices. Then,
by comparing \eref{eqzpadtrap} to \eref{zeqv} we immediately see that for the vortex trapping to generate the same decay rate as the normal-metal trap, the number of vortices in each pad must be equal to:
\be
\bar{N} = \frac{L_{\mathrm{pad}}^2\Gamma_{\mathrm{eff}}}{P}\, .
\ee
Using $P=6.7\times 10^{-2} \mathrm{cm}^2 \;\mathrm{s}^{-1}$ and $L_{\mathrm{pad}}=80~\mu$m \cite{wang}, the number of vortices in each pad would need to be $\bar{N}\simeq 230$. This shows that many vortices are needed to match the efficiency of the normal metal trap.

The cooling magnetic field needed to achieve this vortex number can be estimated to be $B\sim \bar{N} \Phi_0/S_\mathrm{pad} \sim 0.75$~G, well into the regime in which the dissipation caused by the vortices negatively affect the qubit coherence \cite{wang}.
A normal-metal island could also lead to dissipation. However,
solving \eref{eqzpadtrap} for the parameters specified in Fig.~\ref{fig:devb1}, we find a density decay rate $\tau_w^{-1}\approx 1.7\text{ms}^{-1}$ for a metal-covered pad. From Fig.~\ref{fig:devb1} we see that an optimally placed trap of length comparable to $l_0$ can achieve this decay rate, even though the trap area is much smaller than the pad area -- thus, optimal placement can potentially limit the losses due to the normal metal. In contrast, allowing more vortices to enter the qubit (e.g., by fabricating wider wires or increasing the cooling magnetic field) is detrimental; indeed, in \ocite{wang} at higher magnetic fields the qubit coherence time were found to shorten, likely due to vortices in the gap capacitor.

\section{Finite-size trap}
\label{appendix_finite_size_trap}

In this appendix, we treat a 1D system with a finite-size trap to identify the regime in which it can be considered as infinitely
small. Moreover, we describe the crossover from infinitely small to
finite size trap in the strong trapping regime.

\begin{figure}[b]
 \includegraphics[width=0.48\textwidth]{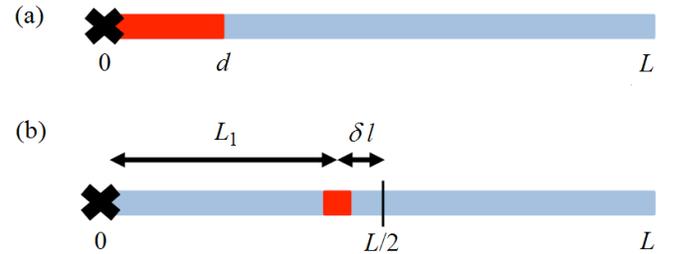}
 \caption{Simplified systems considered in (a) Appendix~\ref{appendix_finite_size_trap} and (b) Appendix~\ref{app:mode_degeneracy}. Blue/light grey denotes superconducting material and red/dark grey the part covered by the normal metal trap. The junction position marked with an X is at the origin of the wire of length $L$.}
 \label{fig:appdiag}
\end{figure}

We consider the 1D diffusion equation (where the spatial coordinate is $0 \le y\le L$)
\be\label{xqp_eq_appb}
\dot{x}_{\text{qp}}=D_{\text{qp}}\nabla^{2}x_{\text{qp}}-\mathcal{A}\left(y\right)\Gamma_{\text{eff}}x_{\text{qp}}.
\ee
We model the trap as a piece of length $d$, starting from
$y=0$, i.e., $\mathcal{A}\left(y\right)=1$ for $y\leq d$ and $0$ otherwise, see Fig.~\ref{fig:appdiag}(a). Since no quasiparticle can leave the ends of the 1D wire, we adopt ``hard wall'' boundary conditions~\cite{notebc}
\be\label{appbbc}
\frac{\partial x_\qp}{\partial y}\Big|_{y=0} = \frac{\partial x_\qp}{\partial y}\Big|_{y=L} = 0\,.
\ee

The time-dependent solution of this problem can be expressed through
the decomposition into eigenmodes, where
\be\label{eigenmodesexp}
x_{\text{qp}}\left(y,t\right)=\sum_{k}e^{\mu_{k}t}\alpha_{k}n_{k}\left(y\right)
\ee
and the eigenmodes fulfill
\be
\mu_{k}n_{k}\left(y\right)=\left[D_{\text{qp}}\vec{\nabla}^{2}-\mathcal{A}\left(y\right)\Gamma_{\text{eff}}\right]n_{k}\left(y\right).
\ee
The eigenvalue problem can be solved with the Ansatz
\be
n_{k}\left(y\right)=\frac{1}{\sqrt{N_{k}}}\left\{ \begin{array}{ll}
\cos\left(\widetilde{k}y\right)\, , & y<d\\
a_{k}\cos\left(ky\right)+b_{k}\sin\left(ky\right) \, , & y>d
\end{array}\right.
\ee
which satisfies the first boundary condition in \eref{appbbc} and we defined
\be
\widetilde{k}=\sqrt{k^{2}-\lambda_{\text{tr}}^{-2}}\, ,
\ee
with $\lambda_\text{tr}$ of \eref{lambdatr} and
$N_{k}$ is a normalization constant. From this Ansatz it follows that $\mu_k = -D_\qp k^2$. Continuity of the function
$n_k$ and its derivative at $y=d$, together with the second condition in \eref{appbbc}, provide an equation for $k$:
\be\label{eq:kcond}
k\tan\left[k\left(L-d\right)\right]=-\widetilde{k}\tan\left(\widetilde{k}d\right).
\ee

While the modes thus defined provide the full time-evolution for all times, we are usually
interested in the lowest mode which dominates the long-time behavior and which we denote with $k_{0}$.
Assuming $k_0\ll \lambda_\text{tr}^{-1}$, we have $\widetilde{k}_0 \approx 1/\lambda_\text{tr}$
and we may approximate \eref{eq:kcond} as
\be\label{eq:kcond2}
l_\text{eff} k=\cot\left[k\left(L-d\right)\right],
\ee
where we defined
\be
l_{\text{eff}}=\lambda_{\text{tr}}\coth\left(\frac{d}{\lambda_{\text{tr}}}\right)\,.
\ee
In the case $d\ll\lambda_{\text{tr}}$ (which implies also
$d\ll L$) \eref{eq:kcond2} becomes
\be
\frac{\lambda_{\text{tr}}^{2}}{d}k=\cot\left(kL\right)\, ,
\ee
which is equivalent to the model where the trap is represented by
a delta function, $\mathcal{A}\left(y\right)\Gamma_{\text{eff}}\rightarrow\gamma_{\text{eff}}\delta\left(y\right)$,
where $\gamma_{\text{eff}}=\Gamma_{\text{eff}}d$ and the trap is
located at $y=0$, see \ocite{Riwar}. Here, in the strong trap limit, $d\gg l_{0}$ with $l_0$ of \eref{eq:swco},
we recover the diffusion-limited lowest mode
\be
k_{0}\approx\frac{\pi}{2L}.
\ee
This solution can be made more general using \eref{eq:kcond2}. Namely, even when $d\gtrsim\lambda_{\text{tr}}$
(i.e., the trap is not small) we may identify the
regime $l_{\text{eff}}\ll L-d$ in which
\be
k_{0}\approx\frac{\pi}{2}\frac{1}{L-d} \, .
\ee
This expression of course coincides with above diffusion-limited solution for
$d\ll L$, and satisfies the initial assumption $k_0 \ll 1/\lambda_\text{tr}$ if $\lambda_\text{tr} \ll L-d$.
Hence we may for instance increase $d$ from $d\ll\lambda_{\text{tr}}\ll L$
to $\lambda_{\text{tr}}\ll d\ll L$, without changing the decay rate,
as long as $l_{\text{eff}}\ll L$. However, the density of the mode
close to the trap changes drastically with increasing $d$.
Indeed, the normalization constant $N_{k}$ in the limit
$l_{\text{eff}},\, d\ll L$ is given by
\be
N_{k}\approx\frac{1}{2}\frac{1}{\lambda_{\text{tr}}^{2}k^{2}}\sinh^{2}\left(\frac{d}{\lambda_{\text{tr}}}\right)
\ee
and hence, the density at the origin for the lowest mode $n_{k_{0}}$ becomes
\be
n_{k_{0}}\left(0\right)\approx\frac{\pi}{\sqrt{2}}\frac{\lambda_{\text{tr}}}{L}\frac{1}{\sinh\left(\frac{d}{\lambda_{\text{tr}}}\right)}
\ee
which, for $d\ll\lambda_{\text{tr}}$ goes as
\be
n_{k_{0}}\left(0\right)\sim\frac{l_{0}}{d} \, ,
\ee
while for $d\gg\lambda_{\text{tr}}$ we find
\be
n_{k_{0}}\left(0\right)\sim\frac{\lambda_{\text{tr}}}{L}e^{-\frac{d}{\lambda_{\text{tr}}}}.
\ee
As we see, by increasing $d$ above the trapping length scale $\lambda_{\text{tr}}$, the density of quasiparticles gets exponentially suppressed near the trap. For a long trap $d\gg \lambda_\text{tr}$, such an exponential suppression takes place also for the steady-state density, as one can verify by reintroducing the generation rate $g$ in the right hand side of \eref{xqp_eq_appb} and solving for the steady-state configuration with $\dot{x}_\qp = 0$.

\section{Quasi-degenerate modes and their observability}
\label{app:mode_degeneracy}

In this appendix, we consider the dependence of the lowest mode on the
trap position. Moreover, we show that close to the optimal position
the lowest and second lowest modes are quasi-degenerate. We finally
comment on the consequences of this quasi-degeneracy on the observability
of the lowest mode.

We take for simplicity a small trap ($d\ll \lambda_\text{tr}$) in a wire of length $L$, placed at an
arbitrary distance $L_1$ from the origin, see Fig.~\ref{fig:appdiag}(b); the quasiparticle density then obeys the diffusion equation (see \cite{Riwar} and App.~\ref{appendix_finite_size_trap})
\be\label{eq:diffappc}
\dot{x}_{\text{qp}}=D_{\text{qp}}\nabla^{2}x_{\text{qp}}-\delta\left(y-L_1\right)\gamma_{\text{eff}}x_{\text{qp}}.
\ee
To solve this equation, we look for eigenmodes $n_{k}$ that must satisfy at $y=L_1$ the condition
\be\label{eq:ylcond}
l_{\text{sat}}\left[\partial_{y}n_{k}\left(L_1+0^+\right)-\partial_{y}n_{k}^{-}\left(L_1-0^+\right)\right]=n_{k}\left(L_1\right)
\ee
with $l_{\text{sat}}= \sqrt{D_\qp t_\text{sat}}$, where $t_\text{sat} = D_{\text{qp}}/\gamma_{\text{eff}}^2$ (this condition follows from the standard procedure of integrating \eref{eq:diffappc} over an infinitesimal interval around $L_1$).
The saturation time $t_\text{sat}$ was introduced in \ocite{Riwar} when studying the quasiparticle dynamics during injection and gives the time scale to reach a steady state. Here we use the related lengths scale $l_\text{sat}$ to have a more compact notation: due to the identity
\be\label{lsat}
\frac{\pi}{2}\frac{l_\text{sat}}{L} = \frac{l_0}{d} \, ,
\ee
this is not an independent parameter in the problem, and the strong (weak) trap condition can be expressed as $l_\text{sat} \ll L$ ($l_\text{sat} \gg L$).

As in the previous Appendix, see \eref{appbbc}, we assume ``hard walls'' on both ends, $\partial_y n_k(0)=\partial_y n_k(L) = 0$; then the eigenmodes are given by
\be
n_{k}=\left\{ \begin{array}{c}
a_{k}\cos\left(ky\right)\text{ for }y<L_1\\
b_{k}\cos\left[k\left(L-y\right)\right]\text{ for }y>L_1
\end{array}\right. \, .
\ee
These modes decay with a rate $1/\tau_k = D_\qp k^2$.
From \eref{eq:ylcond} and continuity of $n_k$, we find the condition for $k$
\be\label{eq:kdcond}
l_{\text{sat}}k\sin\left(kL\right)=\cos\left(kL_1\right)\cos\left[k\left(L-L_1\right)\right].
\ee
For an infinitely strong trap, $l_{\text{sat}}\rightarrow0$, we get
the condition
\be
\cos\left(kL_1\right)\cos\left[k\left(L-L_1\right)\right]=0,
\ee
which provides for the lowest mode either $k=\frac{\pi}{2L_1}$ or $k=\frac{\pi}{2\left(L-L_1\right)}$
depending on whether $L-L_1\gtrless L_1$. Note that the continuity
of the modes at $y=l$ requires
\be\label{eq:kdcont}
a_{k}\cos\left(kL_1\right)=b_{k}\cos\left[k\left(L-L_1\right)\right]
\ee
which means that $b_{k}=0$ for $k=\frac{\pi}{2L_1}$ or likewise $a_{k}=0$
for $k=\frac{\pi}{2\left(L-L_1\right)}$. In other words, the trap effectively separates
the wire into two independent pieces, one to the left and one to right of the trap, with the quasiparticle density of the lowest mode fully suppressed
in the shorter piece.

We define the optimal trap position as the one where the lowest mode decay rate is
the highest. It is easy to see that this is at the degeneracy point, $L_1=L/2$,
where the two modes' decay rates coincide. Note however, that when passing the
degeneracy point by increasing $L_1$ from $L_1<L/2$ to $L_1>L/2$, the
eigenmode function jumps abruptly from being nonzero on the right
hand side to nonzero on the left. Therefore, whether the quasiparticle density
actually decays with the rate defined by the lowest mode, can be strongly affected by the initial conditions.
In order to study this effect, we depart
from the ideal, infinitely strong trap, and take a small
but finite $l_{\text{sat}}$. In addition, we look at a system
close to the degeneracy point, that is, $L_1 = L/2+\delta l$ with
$\delta l\ll L/2$. We first expand \eref{eq:kdcond} for small $\delta l$
\be
l_{\text{sat}}k\sin\left(Lk\right)+\sin^2\left(\frac{Lk}{2}\right) \left(k \delta l\right)^{2}=\cos^{2}\left(\frac{Lk}{2}\right).
\ee
Next, we set $k = \pi/L+\delta k$ and, assuming a strong trap, $l_\text{sat} \ll L$, we expand up to second for $L\delta k \ll 1$ to find
\be
4\pi^{2}\frac{l_{\text{sat}}^{2}+\delta l^{2}}{L^{2}}=\left(L\delta k+\frac{2\pi l_{\text{sat}}}{L}\right)^{2}.
\ee
This results in
\be
\delta k_{\mp}=2\frac{\pi}{L}\left(-\frac{l_{\text{sat}}}{L}\mp\frac{\sqrt{l_{\text{sat}}^{2}+\delta l^{2}}}{L}\right),
\ee
and we see that the degeneracy at $\delta l=0$ has been
lifted by the small parameter $l_{\text{sat}}/L$.

From the continuity condition \eref{eq:kdcont}, we are able to obtain for each mode the ratio between
the (maximal) densities to the left and to the right of the trap
\be
\frac{a_{k_{\mp}}}{b_{k_{\mp}}}=\frac{-\frac{\delta l}{L}-\frac{l_{\text{sat}}}{L}\mp\frac{\sqrt{l_{\text{sat}}^{2}+\delta l^{2}}}{L}}{\frac{\delta l}{L}-\frac{l_{\text{sat}}}{L}\mp\frac{\sqrt{l_{\text{sat}}^{2}+\delta l^{2}}}{L}}.
\ee
In the limit $\delta l\ll l_{\text{sat}}$ this reduces to
\be
\frac{a_{k_{\mp}}}{b_{k_{\mp}}} \simeq \pm 1\, .
\ee
For $l_{\text{sat}}\ll\delta l$, on the other hand, we find
\be
\frac{a_{k_\mp}}{b_{k_\mp}}\approx\pm\left(\frac{2\left|\delta l\right|}{l_{\text{sat}}}\right)^{\pm\text{sign}\delta l},
\ee
which is either very large or very small. This means that in this case the two modes have a very strong asymmetry
in the relative density left and right of the trap.

This asymmetry can affect the measurement of the density decay rate, estimated via a
local measurement of the density in time. Let us suppose that we measure the quasiparticle
density at $y=0$. If $\delta l>0$, the trap is further away
from the detection point and thus the slower mode has a high density
on the detector side; in this case we simply measure the slowest decay rate. On the contrary,
if $\delta l<0$ (with $|\delta l| \gg l_\text{sat}$), the faster mode has most of its density close to
the origin and, depending on the time scale on which we measure,
we may observe the higher decay rate.
Let us suppose we have an initially homogeneous quasiparticle distribution,
so that $b_{k_-} \approx a_{k_+}$.
Due to the asymmetry of the two modes, the initial ($t=0$) ratio $r$ of the densities of the slowest to the faster mode at the origin $y=0$ is
\be
r\left(0\right)\equiv\frac{a_{k_{-}}}{a_{k_{+}}}\approx\frac{l_{\text{sat}}}{2\left|\delta l\right|}\ll1.
\ee
Hence initially, one can observe only the faster mode. As the decay progresses,
this ratio eventually shifts in favor of the lowest mode,
\be
r\left(t\right) = r(0) e^{-D_\qp (k_-^2 - k_+^2)t} \approx \frac{l_{\text{sat}}}{2\left|\delta l\right|}e^{8\frac{\left|\delta l\right|}{L}\frac{t}{t_{D}}},
\ee
where we defined $t_{D}=\pi^{2}D_\qp/L^{2}$.
The time at
which the lowest mode becomes dominant can be estimated by setting $r(t) \sim 1$:
\be
\frac{t}{t_{D}}\sim \frac{L}{8\left|\delta l\right|}\ln\frac{2\left|\delta l\right|}{l_{\text{sat}}} \, ,
\ee
where the right hand side is $\gg1$. Thus, as we see,
the time at which we can observe the decay of the lowest mode
is much longer than $t_{D}$. This long time scale could also affect the choice of optimal trap placement, as we discuss in the next Appendix.

\section{Effective length due to the pad}
\label{app:Effective-length-pad}

In the main text we discuss a device consisting of a long quasi-1D wire (length $L$ and width $W$)
with a square pad (side $L_\text{pad}$) at one end, see Fig~\ref{fig:padtrap}. Here we show that for slow modes,
the presence of the pad can be accounted for by the addition
of an effective length to the original length of the wire.
Indeed, let us assume that the decay times of the modes we are interested in are long compared to the diffusion time $\tau_\text{pad} = L_\text{pad}^2/D_\qp$ across the pad. Then we can take the density in the pad
to be approximately uniform, and this assumption leads to the following boundary condition~\cite{wang}
\begin{equation}
\dot{x}_{\text{qp}}^{\text{wire}}\left(L\right)=-\frac{WD_\qp}{L_{\text{pad}}^{2}}\partial_{y}x_{\text{qp}}^{\text{wire}}\left(L\right)
\label{eq:boundary_with_pad}
\end{equation}
for the density in the wire at the position where it joins the pad.
We now show that this condition leads to a ``hard wall'' boundary condition
for a 1D wire with an effective length which is longer due to the
pad.

We use for $x_{\text{qp}}^{\text{wire}}$ an expansion analogous to \eref{eigenmodesexp}, where each mode in a 1D wire is generally of the form
\be
n_{k}^{\text{wire}}\left(y\right)=a_{k}\cos\left(ky\right)+b_{k}\sin\left(ky\right).
\ee
Substituting such an expansion into \eref{eq:boundary_with_pad}, and remembering that $\mu_k= -D_\qp k^2$, we find
\be\label{eq:bwp2}
a_{k}\left[\widetilde{L}k\cos\left(Lk\right)+\sin\left(Lk\right)\right]=b_{k}\left[\cos\left(Lk\right)-\widetilde{L}k\sin\left(Lk\right)\right]
\ee
where we use the notation $\widetilde{L}=L_{\text{pad}}^{2}/W$. Defining the
effective length addition for mode $k$ as
\be\label{eq:Lepk}
L^{\text{eff}}_\text{pad}\left(k\right)=\frac{1}{k}\arcsin\left(\frac{\widetilde{L}k}{\sqrt{1+\widetilde{L}^{2}k^{2}}}\right)
\ee
we rewrite \eref{eq:bwp2} as
\be
\tan\left[\left(L+L^{\text{eff}}_\text{pad} \left(k\right)\right)k\right]=\frac{b_{k}}{a_{k}},
\ee
which indeed has the same form of the ``hard wall'' boundary condition for a wire of length $L+L^{\text{eff}}_\text{pad}$.
For the limiting case $\widetilde{L}k\ll1$, we
find that $L^{\text{eff}}_\text{pad}\approx\widetilde{L}$ and hence, in this
case, the effective total system length is $L+L_{\text{pad}}^{2}/W$.

\section{Effective length due to the gap capacitor}
\label{app:Effective-length-wings}

Similarly to Appendix~\ref{app:Effective-length-pad}, we show here that the gap capacitor provides an effective extension of the central wire.
For this purpose, we add to one end of
the wire of length $L$, two perpendicular wires, each of length $L_c$
and width $W_c$, cf. Fig.~\ref{fig:padtrap} in the main text. Current conservation at the junction between the three wires
provides the condition
\begin{equation}
W\left.\partial_{y}x_{\text{qp}}^{\text{wire}}\right|_{y=L}=-2W_c\left.\partial_{x}x_{\text{qp}}^{c}\right|_{x=L_{c}}.
\label{eq:boundary_side_wings}
\end{equation}
Here, we assumed that the wire (gap capacitor) density is constant in the $x$-
($y$-) direction. The eigenmodes of wire and capacitor are of the form
\begin{eqnarray*}
n_{k}^{\text{wire}}\left(y\right) & = & a_{k}\cos\left(ky\right)+b_{k}\sin\left(ky\right),\\
n_{k}^{c}\left(x\right) & = & c_{k}\cos\left(kx\right).
\end{eqnarray*}
Substituting this Ansatz into \eref{eq:boundary_side_wings} and
requiring continuity at junction, we find
the condition
\begin{eqnarray*}
 & a_{k}\left[\sin\left(Lk\right)\cos\left(L_{c}k\right)+2\frac{W_c}{W}\sin\left(L_{c}k\right)\cos\left(Lk\right)\right]\\
 & =b_{k}\left[\cos\left(Lk\right)\cos\left(L_{c}k\right)-2\frac{W_c}{W}\sin\left(L_{c}k\right)\sin\left(Lk\right)\right].
\end{eqnarray*}
Defining the effective length addition due to the capacitor as
\be\label{eq:Leck}
L_c^{\text{eff}}\left(k\right)=\frac{1}{k}\arcsin\left(\frac{2\frac{W_c}{W}\tan\left(L_{c}k\right)}{\sqrt{1+4\frac{W_c^{2}}{W^{2}}\tan^{2}\left(L_{c}k\right)}}\right),
\ee
we find the effective hard wall boundary condition
\be
\frac{b_{k}}{a_{k}}=\tan\left[\left(L+L_c^{\text{eff}}\left(k\right)\right)k\right].
\ee
Note that for $2W_c/W=1$, the capacitor represents simply
a direct extension to the wire with $L_c^{\text{eff}}=L_{c}$.
For $L_{c}k\ll1$ and $2W_c/W \ll 1/(L_c k)$,
we may approximate $L_c^{\text{eff}}\approx 2\frac{W_c}{W}L_{c}$.

\section{Traps in the Xmon geometry}
\label{app:Xmon}

\begin{figure}[tb]
\parbox{0.21\textwidth}{
\qquad \includegraphics[width=0.19\textwidth]{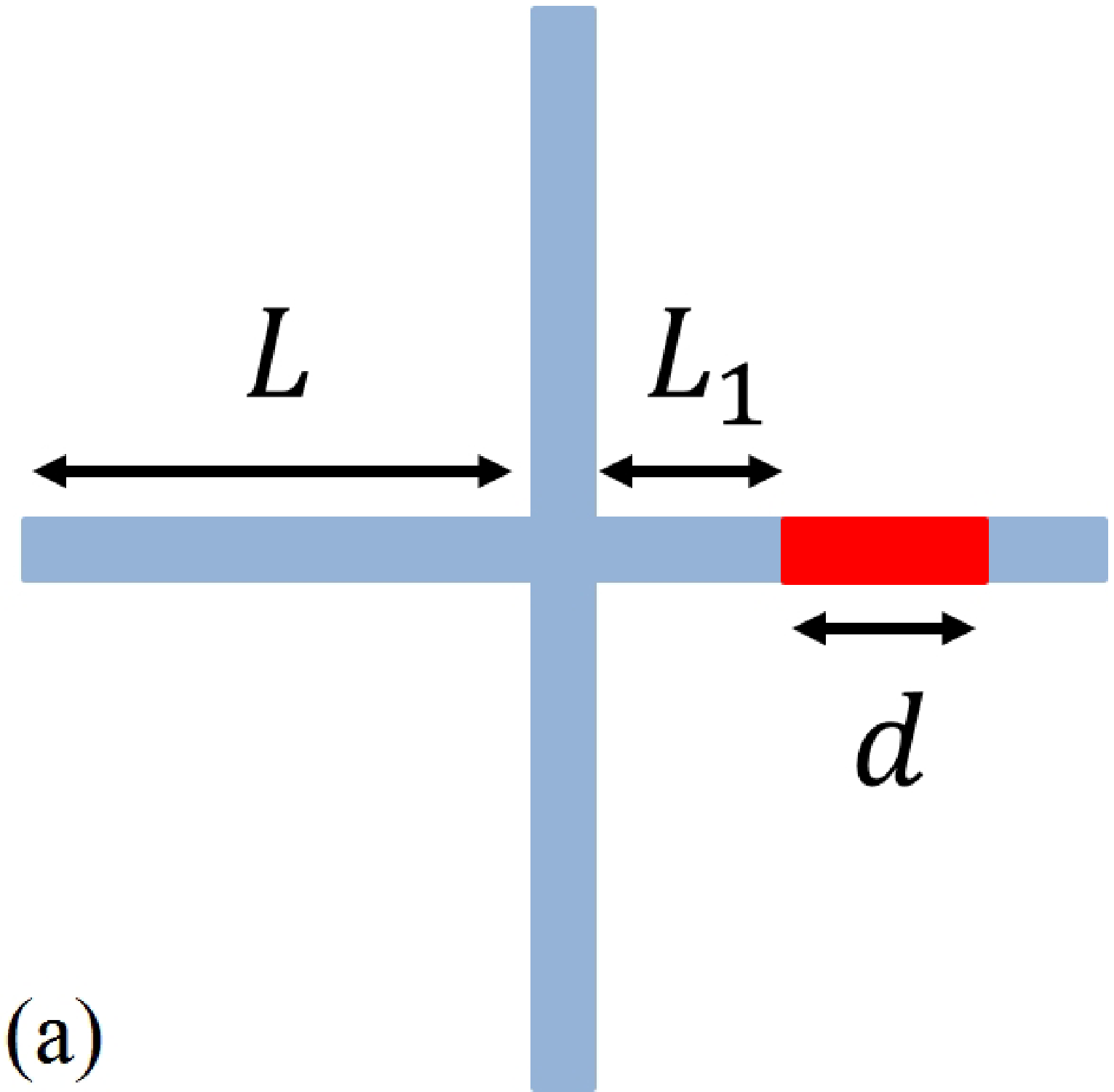}} \qquad \,
\parbox{0.21\textwidth}{
\includegraphics[width=0.19\textwidth]{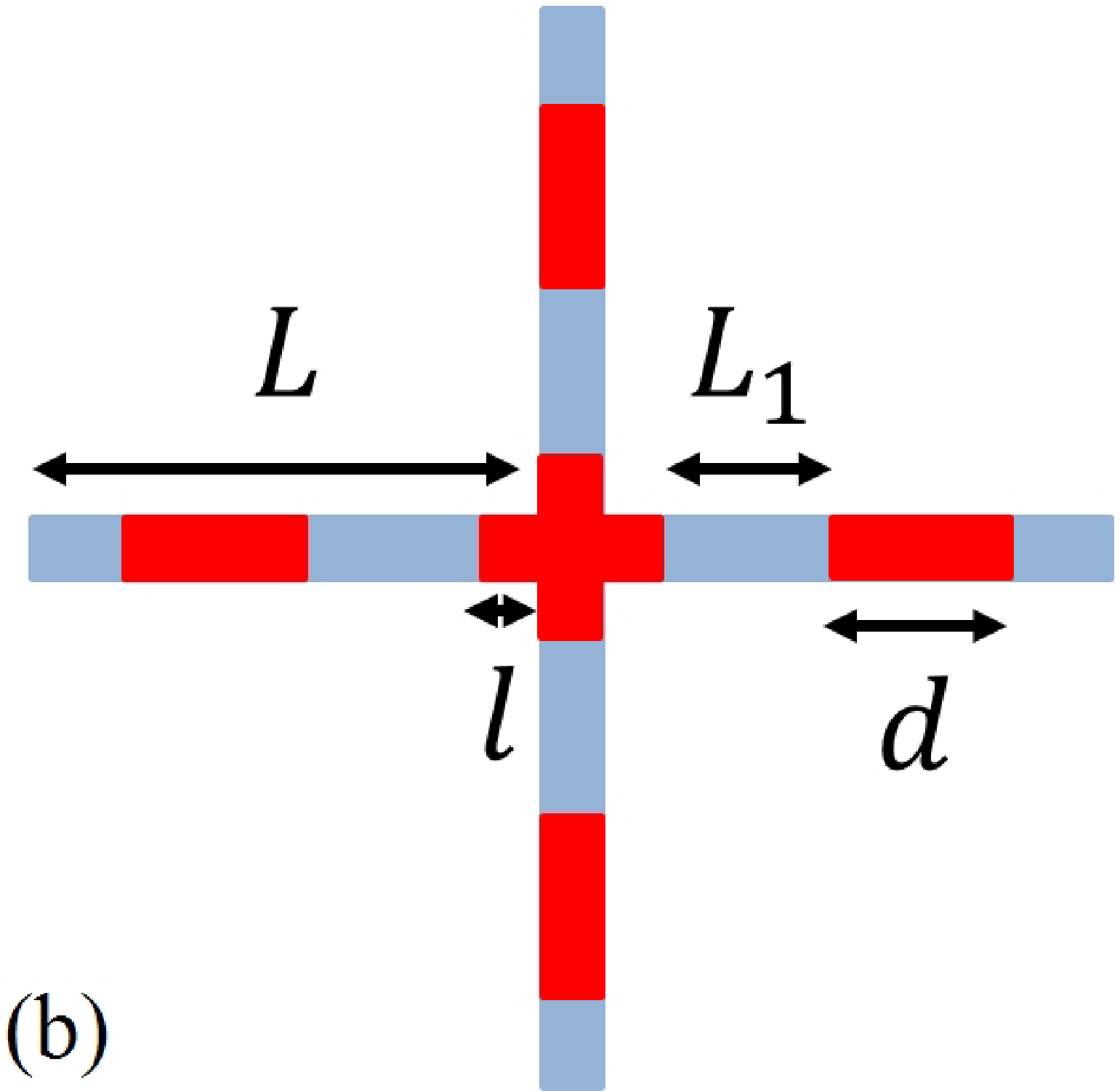}}
\\ \vspace{2mm}
\includegraphics[width=0.47\textwidth]{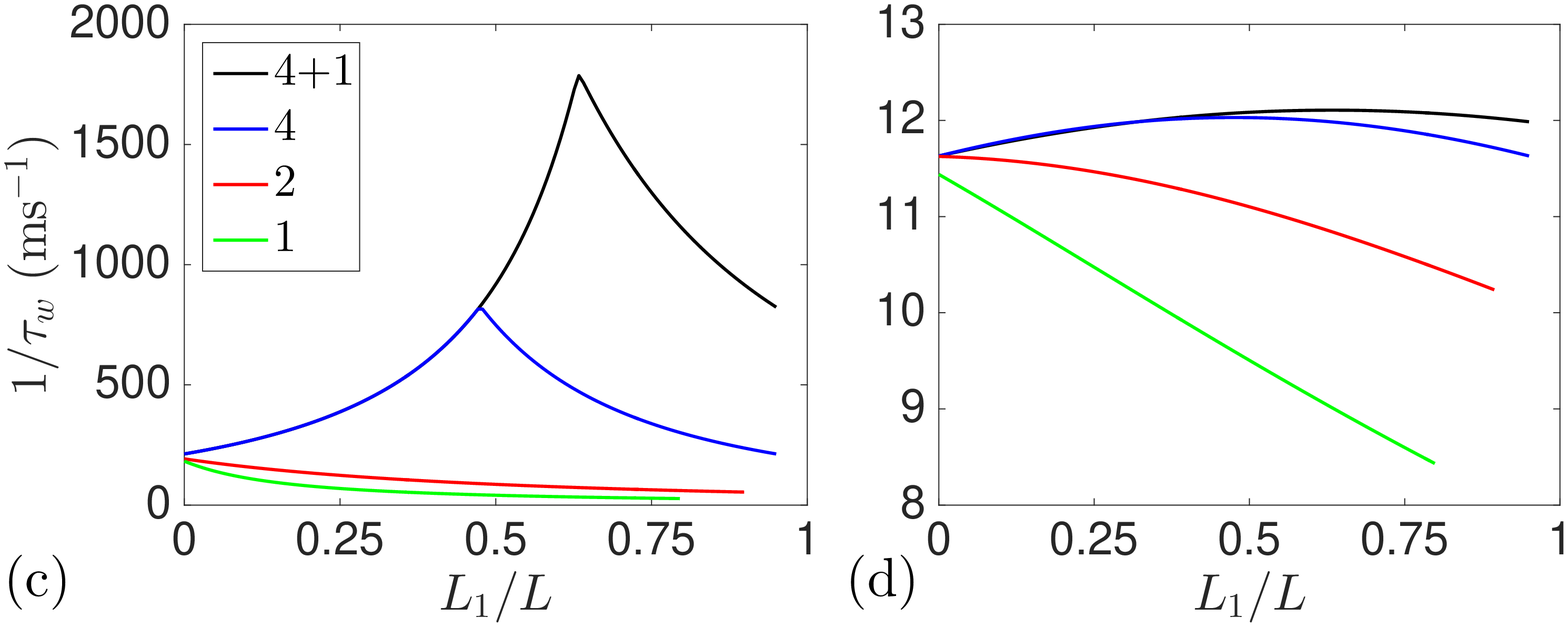}
\caption{(a) The single trap geometry. The trap is at distance $L_1$ from the center, and has size $d$. The $2$ trap configuration follows from this setup, by adding a trap of the same size on the left branch, with the same distance $L_1$ from the center. (b) The $4+1$ trap geometry. All traps on the individual arms have the same distance $L_1$ from the center. The total trap size is $d'=4l+4d$. The $4$-trap geometry follows from this one by removing the middle cross-like trap. (c) The resulting decay rate $\tau_w^{-1}$ as a function of $L_1$, for (bottom to top) $1$, $2$, $4$, and $4+1$ traps. The parameters are $L=150~\mu$m and $\lambda_\text{tr}=2~\mu$m; the total trap length is $30~\mu$m in all cases. (d) The decay rate as in (c), but with a realistic value $\lambda_\text{tr}=86.3\,\mu$m for the trapping length.}\label{four_arm_geometry}
\end{figure}

In this Appendix we further explore the role of device geometry by studying the optimal placement of traps in the so-called Xmon qubit of Ref.~\cite{barends}. We thus consider a four-arm geometry with symmetric arm lengths, see Fig.~\ref{four_arm_geometry}. Clearly, the optimal position for a single trap is at the center of the device; however, having two or three traps cannot lead to large improvement in the decay rate with respect to one trap, because the  diffusion time cannot be shortened in all arms. Therefore, we need at least four traps, one in each arm, to improve $\tau_w^{-1}$. A fifth should again be placed at the center, rather than in the arms. In fact, by generalizing the argument given at the beginning of Sec.~\ref{sec:multitrap}, we find that the decay rate scales as $(N_\text{tr}/2)^2$ if $N_\text{tr}$ is multiple of 4, and as $[(N_\text{tr}+1)/2]^2$ if $N_{tr} = 4n+1$, $n=0,\,1,\ldots$ While in both cases the scaling is less favorable than the $N_\text{tr}^2$ one for a single wire, we see that for a small number of traps the configuration with the additional trap at the center gives a larger increase in the trapping rate.

To validate the above considerations, we solve the diffusion equation in the geometry obtained by simply joining four equally long 1D wires of length $L$.
We consider $1$, $2$, $4$ and $4+1$ traps, all with the same total area, placed symmetrically as depicted in Fig.~\ref{four_arm_geometry}(a) and (b).
We show the resulting decay rate $\tau_w^{-1}$ in Fig.~\ref{four_arm_geometry}(c) for a strong trap, obtained by assuming $\lambda_\text{tr} \ll L$. Comparing to
the single-trap case, we find the expected improvement by a factor of $4$ ($9$) for $4$ ($4+1$) traps. However, in an actual device the length is $L\approx 150\,\mu$m, which
is not much larger than the estimate $\lambda_\text{tr}\approx86.3\,\mu\text{m}$ and gives, using \eref{eq:swco}, a trap size $l_0 \simeq 78\,\mu$m for the cross-over from weak
to strong (diffusion-limited) trap. Therefore, in practice the traps are in the weak regime and their placement does not affect much the decay rate, see Fig.~\ref{four_arm_geometry}(d). This points to the need for stronger traps (with shorter $\lambda_\text{tr}$) for effective trapping in small devices. Alternatively, one could use traps in the ground plane surrounding the small device, to confine most quasiparticles away from it, see \cite{mcdvav}.


\begin{thebibliography}{30}
\bibitem{noterecomb}When calculating decay rates, a finite $s_b$ can be simply added to those obtained for $s_b=0$; for typical parameters, this addition is small. Equivalently, we note that the length scale $\sqrt{D_\qp/s_b}$ is much longer than all other length scales in the problem, which justifies neglecting $s_b$. We also neglect recombination, see \ocite{Riwar}.
\bibitem{notebc}The validity of the boundary condition at $y=0$ can be justified by imagining a short piece of uncovered superconductor extending to the left of the trap, $-\epsilon < y<0$. Then clearly we have $\partial x_\qp/\partial y|_{y=-\epsilon} = 0$; taking the limit $\epsilon\to 0$ and since $x_\qp$ and its derivative are continuous, we arrive at \eref{appbbc}. A particular case is that of a point-like trap with $d\to0$ and a finite, fixed value of $\gamma_\text{eff}=d\Gamma_\text{eff}$; then one can either follow the procedure just described, or equivalently, use the boundary condition $\partial x_\qp/\partial y |_{y=0} = x_\qp(0) \gamma_\text{eff}/D_\qp$, see \eref{eq:ylcond}.
\end{thebibliography}
\end{document}